\documentclass[letterpaper,inpress]{jdsart}

\pubmonth{July}
\pubyear{2022}
\volume{xx}
\issue{xx}
\doi{0000}

\usepackage[utf8]{inputenc}

\usepackage{cleveref}
\usepackage{tabu,booktabs}
\usepackage{amsfonts,amsmath,amssymb,amsthm}

\begin{document}
\begin{frontmatter}

\title{Demonstrative Evidence and the Use of Algorithms in Jury Trials}
\runtitle{Algorithms in Jury Trials}

\author[1]{
  \inits{R.}
  \fnms{Rachel}
  \snm{Rogers}  \thanksref{1}  \ead{rachel.rogers@huskers.unl.edu}}
\author[1]{
  \inits{S.}
  \fnms{Susan}
  \snm{VanderPlas}}

\thankstext[type=corresp,id=1]{Corresponding author}
\address[1]{Department of Statistics, 
  \institution{University of Nebraska-Lincoln}, \cny{United States of America}}

\begin{abstract}
We investigate how the use of bullet comparison algorithms and demonstrative evidence may affect juror perceptions of reliability, credibility, and understanding of expert witnesses and presented evidence. The use of statistical methods in forensic science is motivated by a lack of scientific validity and error rate issues present in many forensic analysis methods. We explore what our study says about how this type of forensic evidence is perceived in the courtroom -- where individuals unfamiliar with advanced statistical methods are asked to evaluate results in order to assess guilt. In the course of our initial study, we found that individuals overwhelmingly provided high Likert scale ratings in reliability, credibility, and scientificity regardless of experimental condition. This discovery of scale compression - where responses are limited to a few values on a larger scale, despite experimental manipulations - limits statistical modeling but provides opportunities for new experimental manipulations which may improve future studies in this area.
\end{abstract}


\end{frontmatter}

\hypertarget{supplementary-material}{%
\section*{Supplementary Material}\label{supplementary-material}}
\addcontentsline{toc}{section}{Supplementary Material}

The Supplementary Material includes: (1) Statistical models and additional graphs for study questions; (2) Code for the creation of the survey app; (3) Survey data and testimony outline; (4) Source files for paper.

\hypertarget{introduction}{%
\section{Introduction}\label{introduction}}

The prevailing belief in bullet comparison in the forensic sciences is that guns can leave individualizing marks on bullets when fired, which can be used to identify the gun \citep{pcast2016}.
Current bullet matching methods rely on a subjective visual comparison of bullets completed by a forensic scientist in order to reach a conclusion \citep{nrc2009}.
In order to improve upon the bullet matching method with increased scientific validity, \citet{pcast2016} urged the development of objective methods of analysis.
These reports have spurred increased research, development, and assessment of statistical matching methods for firearms analysis, including \citep{hare2017, vanderplas2020, songDevelopmentBallisticsIdentification2012}.

As these algorithms are developed and validated, it becomes more important to understand how they may impact the evidentiary process - how will jurors react to algorithms used to match bullets?

In this factorial study, we examine the effect of algorithm use and demonstrative evidence (photos and data visualizations) in jurors' perception of examiner testimony.
We assess the perception of the strength of evidence, guilt or innocence, examiner credibility, and the reliability and scientific validity of firearms examination.
This study is intended to lay the groundwork for the use of algorithmic firearms comparisons in court.
When presented with the same evidence, it is important to know whether or not images affect potential jurors' views on the reliability or credibility of the witness and the evidence that they present.

\hypertarget{bullet-matching-algorithm}{%
\subsection{Bullet Matching Algorithm}\label{bullet-matching-algorithm}}

\begin{figure}
\includegraphics[width=0.49\linewidth]{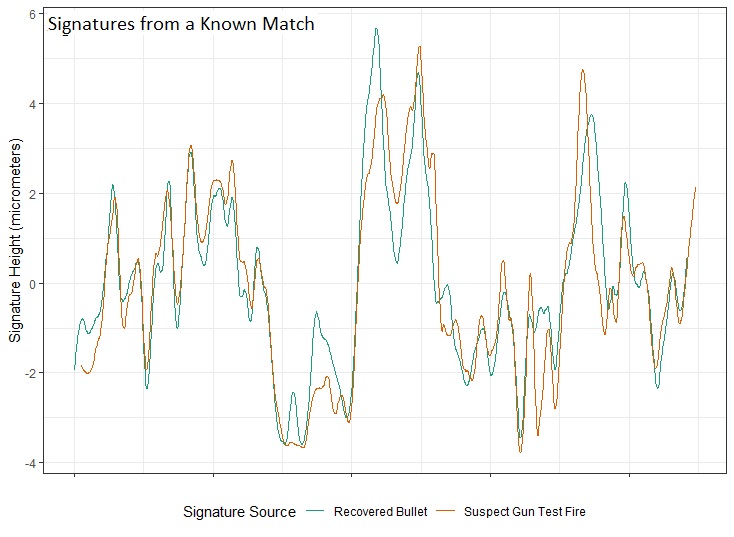} \includegraphics[width=0.49\linewidth]{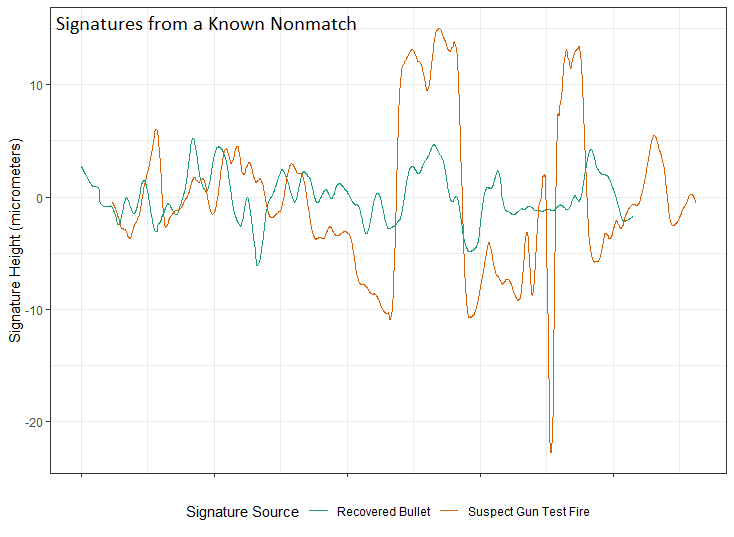} \caption{Bullet signatures for two lands. The left image indicates two matching lands, while the right image indicates two non-matching lands.}\label{fig:signature}
\end{figure}

\begin{figure}
\includegraphics[width=0.49\linewidth]{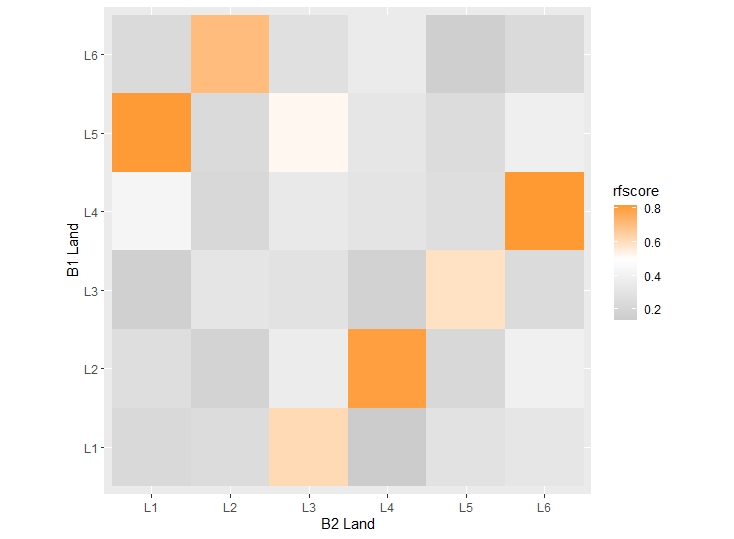} \includegraphics[width=0.49\linewidth]{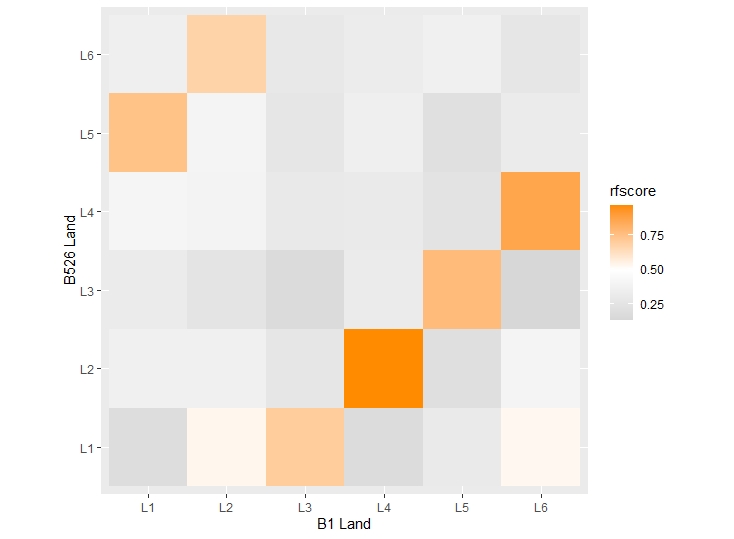} \includegraphics[width=0.49\linewidth]{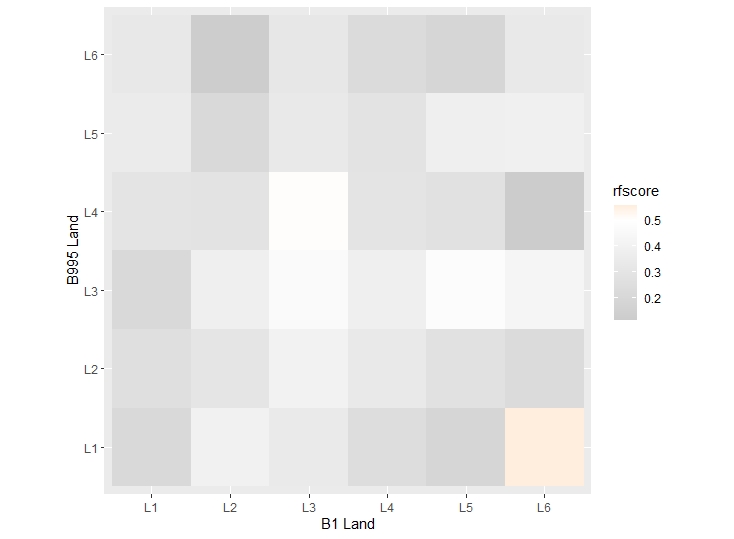} \includegraphics[width=0.49\linewidth]{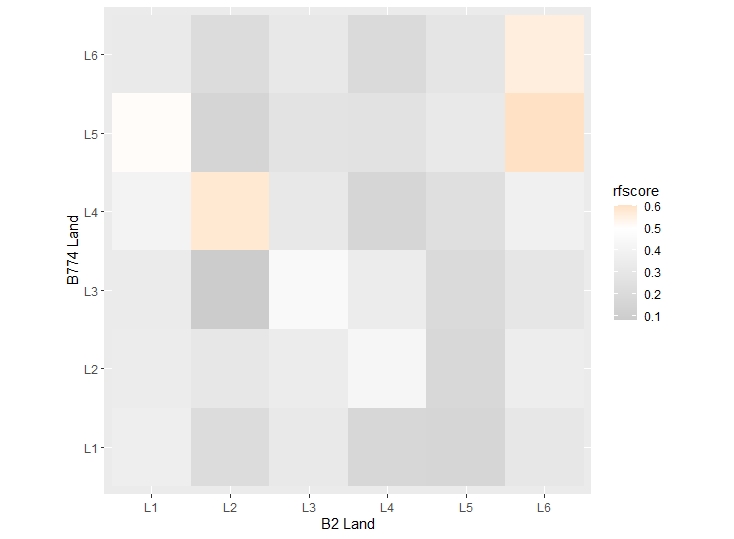} \caption{Comparison grids demonstrating various bullet comparisons. The top two images were from the same source, and were used as the test fire (left) and the algorithmic identification (right) in the sample testimony. The bottom two images are from different sources, and were used as the algorithmic elimination (left) and inconclusive (right) in the sample testimony.}\label{fig:comparisongrid}
\end{figure}

The motivating idea of bullet matching is that there are unique scratches in a gun barrel resulting from the rifling process that make it possible to identify whether two bullets were fired from the same gun \citep{nrc2009}.
This barrel rifling creates a spiral of raised areas, known as lands, and indented areas, known as grooves; these lands leave scratch marks on the bullet that can then be compared using an algorithm \citep{hare2017}.
The transition period between lands and grooves of the bullet are known as shoulders \citep{hare2017}.

Firearms examiners visually examine the lands of bullets using a comparison microscope, which allows two bullets to be directly compared in the same viewfinder.
If there is sufficient similarity (which is a subjective conclusion), the examiner will make an \textbf{identification}, suggesting that the two bullets are from the same source (SS) - they were fired from the same gun.
If there is sufficient dissimilarity (again, a subjective decision), the examiner will make an \textbf{elimination}, concluding that the bullets were not fired from the same gun.
If there is insufficient evidence in either direction (at least in theory, see \citep{inconclusives}), then the examiner will return an \textbf{inconclusive} decision.
The bullet matching algorithm developed by \citet{hare2017} follows these steps: first, a 3D scan is taken of each bullet land, a stable cross-section is extracted, and shoulders (edges) are removed; then a smoothing function is applied twice in order to extract a representative profile, called the signature, which can be compared to signatures from other bullets (as shown in \Cref{fig:signature}); finally, traits derived from the signature are combined using a random forest classifier to produce a match score for each land, ranging in value from 0 to 1, where 0 is indicative of a different-source pair and 1 is indicative of a same-source pair.
There are multiple lands per bullet, resulting in a grid of land-to-land match scores when comparing two bullets, as shown in \Cref{fig:comparisongrid}.
Lands are ordered sequentially, so that bullets from the same source should produce high land-to-land match scores as the land number progresses \citep{vanderplas2020}.
For example, in the top left grid of \Cref{fig:comparisongrid}, Land 5 (L5) of B1 corresponds to Land 1 (L1) of B2, as can be seen by the higher random forest score.
As the bullets are rotated, the lands continue to correspond (Land 6 of B1 to Land 2 of B2, and so forth) in a diagonal - indicating that the bullets match.

\begin{table}
\caption{\label{tab:matchtab}Match Score and Examiner Conclusion Language Used for Algorithm Evidence in the Sample Testimony}
\centering
\begin{tabu}{p{0.2\textwidth}p{0.2\textwidth}p{0.6\textwidth}}
\toprule
Bullet Comparison & Match Score & Language\\
\midrule
Test Fire & 0.976 & \\\\
Identification & 0.989 & The match score indicates that there is substantial similarity between the two bullets, which suggests that they were most likely fired from the same barrel.\\\\
Elimination & 0.034 & The match score indicates that there is significant disagreement between the two bullets.\\\\
Inconclusive & 0.34 & The match score indicates that there is not sufficient agreement between the two bullets, which suggests that the results are inconclusive.\\
\bottomrule
\end{tabu}
\end{table}

These individual scores for the maximum correspondence between the two bullets are averaged to create a bullet-level match score \citep{vanderplas2020}.
The match scores reported in the sample testimony for the corresponding image in \Cref{fig:comparisongrid} are shown in \Cref{tab:matchtab}.
\citet{vanderplas2020} validated the algorithm for use with guns outside models used in the training data.
This is a critical step required before the algorithm could be used in forensic labs.
For the algorithm to be used in practice, though, we must understand how jurors without statistical expertise understand and interpret the results from the algorithm when these results are presented during testimony.
This problem of jury interpretation of statistical results has been encountered before, in disciplines such as DNA and fingerprints \citep{koehler2001, garrett2018}.

\hypertarget{explainable-machine-learning-in-courts}{%
\subsection{Explainable Machine Learning in Courts}\label{explainable-machine-learning-in-courts}}

\citet{swofford2022} interviewed judges, lawyers, scientists, and researchers in order to assess their feelings about the use of statistical methods and probabilistic language in court.
They found that some individuals were concerned about jurors' ability to understand and properly interpret probabilistic language; some suggested that a mix of probabilistic language and match language may be more beneficial than strictly using one or the other.

FRStat (Friction Ridge Statistics), an analysis program which assigns statistical values to fingerprint analyses, uses likelihood ratios to characterize strength of evidence, with phrasing like ``The probability of observing this amount of correspondence is approximately {[}X{]} times greater when the impressions are made by the same source rather than by different sources'' \citep{DFSCLPInformation2018}.
\citet{garrett2018} found that jurors did not provide significantly different likelihoods that the individual was the source of the prints when they were presented with a wide range of FRStat likelihood results, ranging from values of 10 times greater to 100,000 times greater.

In another study, \citet{koehler2001} investigated the perception of probabilities and frequencies in the case of DNA.
They found that individuals were more likely to believe the subject was the source of the DNA when the same number was presented as a probability rather than a frequency.
They also asked individuals to identify the number of people that would match DNA for their given match proportion in a population of 500,000; 60.7\% of those given a frequency and 42.1\% of those given a probability were able to correctly identify the number.
These examples illustrate that there is (justifiable) concern for how statistical methods and results may be presented and interpreted in the courtroom.
In the application of \citet{hare2017}'s algorithm, the correspondence between bullets is described as a match score produced by a random forest, with values between 0 and 1.
This machine learning application may add another hurdle in juror understanding.

\hypertarget{demonstrative-evidence}{%
\subsection{Demonstrative Evidence}\label{demonstrative-evidence}}

Demonstrative evidence, such as images, can serve as an aid in explaining results and methods used in the forensic sciences.
However, there is the potential that the use of images can be biasing.
In a study conducted by \citet{cardwell2016}, researchers found that topically related images may make a scenario more believable, even if the images provide no additional evidentiary value.
Individuals were asked to `give' food to animals represented as words, then were later presented with animals (either as words or accompanied with an image) and asked to identify if they had given food to the animal \citep{cardwell2016}.
Participants were more likely to believe that they had given food to the animal if it was accompanied with an image, regardless of whether or not they had actually given food \citep{cardwell2016}.
Alternatively, In a series of studies asking jurors to evaluate the mental state of the defendant at the time of the crime, \citet{schweitzer2011} found no effect of non-informative neuroimages on jurors' judgements.

In the courtroom, \citet{kellermann} describes the use of non probative images to elicit responses from juries in the form of ``truthiness'' (feelings that a statement is true) or ``falsiness'' (feelings that a statement is false), without introducing additional information through the images.
As statistical graphics can improve our ability to understand data and model results, it is possible that the use of explanatory images may increase jurors' ability to understand the use of algorithms for evaluating forensic evidence.
These graphics differ from those in \citet{cardwell2016} as they are directly showing evidence that is also being presented and explained verbally.
Despite these differences, it is still possible that these images may influence potential jurors' perceptions of the speaker, or their feelings of ``truthiness'' in the case.

\hypertarget{methods}{%
\section{Methods}\label{methods}}

\hypertarget{participants}{%
\subsection{Participants}\label{participants}}

Participants were recruited using Prolific, an online platform for scientific research.
Prolific offers researchers the ability to obtain a representative sample of participants from a specific region (in this case, the United States) across age, race, and gender.
Individuals were additionally asked to self-screen for jury eligibility (no past felony convictions, over the age of majority, not emergency response personnel, etc.).
Participants were paid \$8.40 for their participation in the study and completed the study with a median response time of around 18 minutes.

\hypertarget{online-jury-studies}{%
\subsection{Online Jury Studies}\label{online-jury-studies}}

While every attempt was made to use a representative sample in this study, there are certain unavoidable biases that are present in online jury research \citep{garrett2020}, particularly when transcripts are used in place of videos.
Individuals who participate in Prolific surveys may not be representative of eligible jurors in the United States.
These individuals also do not undergo the jury selection process, and the jury selection process does not provide a representative sample of individuals in the United States \citep{abramson2018}.
In order to provide a study of reasonable length, testimony was limited to the relevant firearms evidence, excluding other witnesses and evidence that may have been presented in a real trial.
Finally, jurors are unable to deliberate as a group; this may result in different conclusions than would be reached under the group dynamic present in deliberation \citep{bornsteinJuryDecisionMaking2011, maccounAsymmetricInfluenceMock1988}.
While acknowledging the limitations of this study format, it is important to note that research on actual jury pools is nearly impossible to conduct for many different logistical and practical reasons, including privacy, cost, and access; even if these barriers were overcome in one jurisdiction, the results from that jurisdiction would not be nationally representative or even representative beyond the sampling area.
Thus, online jury studies are an important tool to understand the effect of different manipulations of courtroom procedures, instructions, and admissible evidence.

\hypertarget{design}{%
\subsection{Design}\label{design}}

In order to assess the effect of evidence presentation and the use of algorithms on how jurors evaluated firearms evidence, we developed a factorial experiment, manipulating the examiner's conclusion (identification, inconclusive, or elimination), whether algorithm testimony was included, and whether testimony included demonstrative evidence (pictures and charts), for a \(3\times 2\times 2\) factorial experiment.
Participants were randomly assigned to one of the twelve experimental conditions.

\begin{table}
\caption{\label{tab:contab}Conclusion Language Used in the Firearms Examiner Testimony}
\centering
\begin{tabu} to \linewidth {l>{\raggedright}X}
\toprule
Conclusion & Language\\
\midrule
Identification & I found that there were sufficient individualizing characteristics to make an identification, that is, that the two bullets were fired from the same barrel.\\\\
Inconclusive & I found that the class characteristics of the two bullets were the same, but there was not sufficient agreement among the individual characteristics. My comparison was inconclusive.\\\\
Elimination & I found that there was significant disagreement in individual characteristics.\\
\bottomrule
\end{tabu}
\end{table}

Participants were presented with a trial scenario either with or without the use of a bullet matching algorithm.
When the algorithm was absent, participants were provided with testimony from a bullet comparison conducted by a firearms examiner, including the comparison of two test fires from the questioned gun in order to establish a baseline.
The wording of the examiner's conclusion is shown in \Cref{tab:contab}.
When the algorithm was present, participants read the same firearms examiner testimony, with an additional algorithmic comparison of the bullets (which supported the firearms examiner's conclusion).
In the transcript, the firearms examiner explained their training in the bullet matching algorithm as well as that the algorithm produces a score between 0 and 1 (0.976 for the test fires from the gun, 0.034 for the elimination condition, 0.34 for the inconclusive condition, and 0.989 for the identification condition).
The transcript then shows the examiner's interpretation of the algorithm results in reference to their own conclusion, as shown in \Cref{tab:matchtab}.
Following the firearms examiner's testimony, the transcript provided testimony from an individual involved in the development of the algorithm, describing the algorithm's process and limitations.
This is consistent with the way DNA comparison algorithms were presented before these algorithms were ubiquitous: a representative from the company providing the algorithm would testify about its development.
Similar situations often arise when investigators make use of algorithms for triangulating a phone's location or linking posts made under different accounts to the same person.
Once an algorithm's use becomes commonplace, the algorithm expert is often not required to testify, but our goal was to assess the initial stage of the use of a bullet matching algorithm in practice.

In demonstrative evidence conditions, images demonstrating barrel rifling \citep{105mmTank2005}, a fired bullet \citep{gremi}, and a comparison microscope with striation marks were included in the testimony.
When both demonstrative evidence and the algorithm were used, the testimony also included an image of the land-to-land comparison grids.
Grids generated by the algorithm were shown for test fires, which were fired from the same gun, and should result in an identification, and for the questioned bullet comparison, reflecting the conclusion of the firearms examiner, as shown in \Cref{fig:comparisongrid}.
The testimony of the algorithm expert included images of the initial cross section scan with lines indicating shoulder removal \citep{hare2017}, and signature comparison for matching and non-matching bullet lands, show in \Cref{fig:signature}.
Presented algorithm match scores and demonstrative evidence were derived from bullet scans which were evaluated by trained firearms examiners as part of an unpublished study, to ensure that the additional information presented was properly calibrated to the design scenario.

\hypertarget{study-format}{%
\subsection{Study Format}\label{study-format}}

\begin{figure}
\centering
\includegraphics[width=\textwidth]{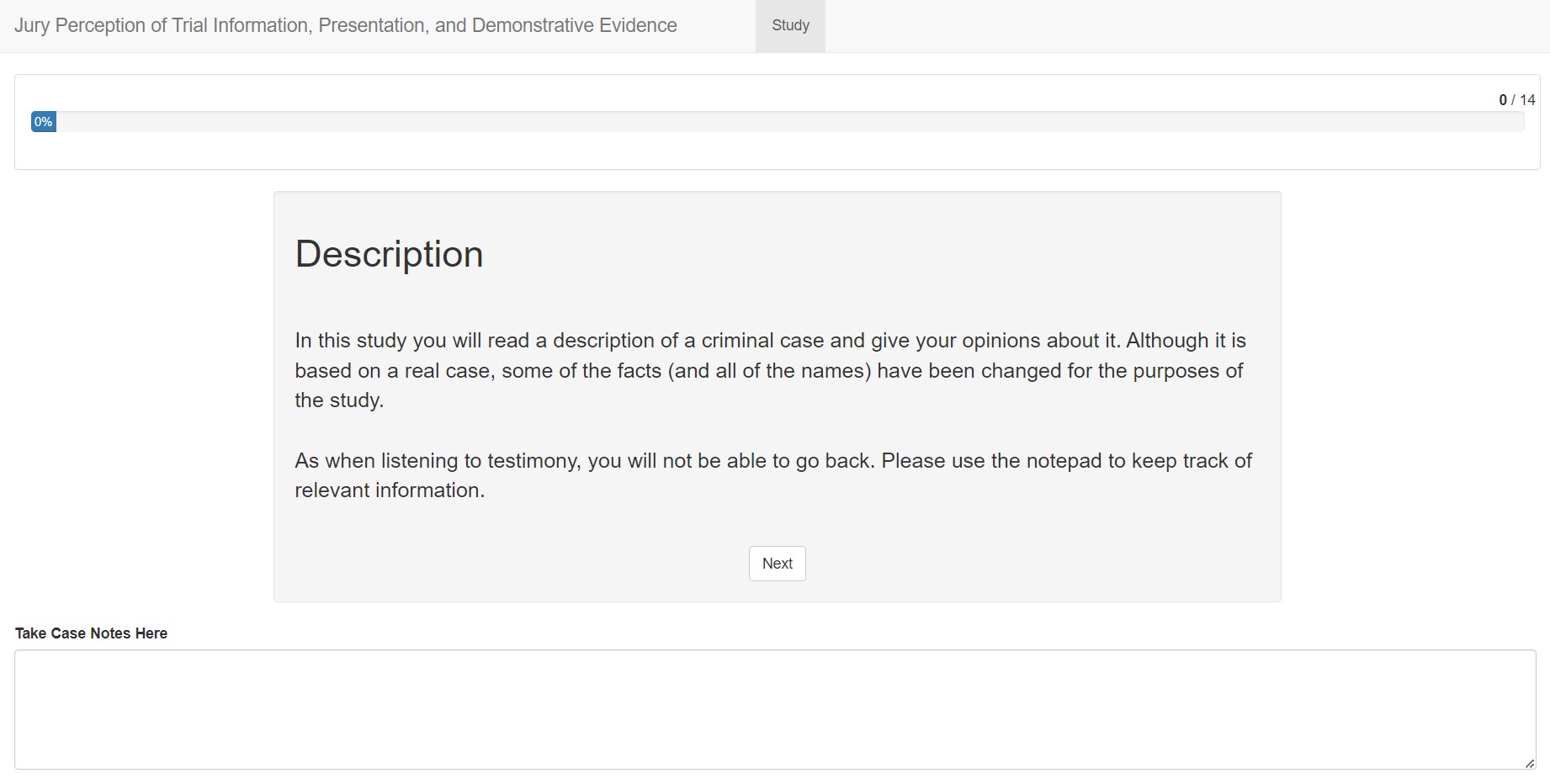}\caption{Screenshot of the study description on the first page of the Shiny app.}\label{fig:screenshot}
\end{figure}

Participants were asked to read a short excerpt of court testimony with regards to an attempted robbery -- a scenario based on \citet{garrett2020}.
In this case, the only evidence linking the defendant, Richard Cole, to the crime scene is a comparison between a gun found in the car and a bullet recovered from the crime scene.
The transcripts were based on testimony given in real trials, as were our edits creating language about algorithms and quantitative evidence.
In order to facilitate participants' recall and provide insight into the portions of the testimony participants found to be important, participants were provided with a way to take notes throughout the presentation of testimony, as jurors are allowed to take notes during the trial for use as a memory aid during deliberation.
At the end of the transcript(s), participants were asked to rate their impression of the evidence presented, as well as their impression of the expert witnesses using Likert scales.
The survey was created using R Shiny \citep{shiny}.
\Cref{fig:screenshot} depicts a screenshot of the study description.

\hypertarget{results}{%
\section{Results}\label{results}}

Five hundred and ninety-one participants completed the survey; of these, 569 correctly answered both attention check questions, identifying the caliber of the gun used in the crime and selecting a response indicated by the question text.
The attention check questions were present to ensure that online participants were reading the testimony as well as the questions before selecting answers.
All 569 participants correctly answering both attention check questions were included in the analyses reported below.
The number of participants for each of the 12 conditions is shown in \Cref{tab:contottab}, and demographic information is shown in \Cref{tab:ethnicitytab}.

\begin{table}

\caption{\label{tab:contottab}Number of Participants per Condition}
\centering
\begin{tabular}[t]{l|l|r|r|r}
\hline
Algorithm & Picture & Elimination & Inconcl. & Identification\\
\hline
No & No & 49 & 54 & 55\\
\hline
No & Yes & 49 & 51 & 52\\
\hline
Yes & No & 50 & 36 & 32\\
\hline
Yes & Yes & 48 & 50 & 43\\
\hline
\end{tabular}
\end{table}

\begin{table}

\caption{\label{tab:ethnicitytab}Number of Participants based on Ethnicity}
\centering
\begin{tabular}[t]{l|r}
\hline
Ethnicity & Count\\
\hline
Asian & 36\\
\hline
Black & 75\\
\hline
Mixed & 12\\
\hline
Other & 10\\
\hline
Unknown & 4\\
\hline
White & 463\\
\hline
\end{tabular}
\end{table}

The average age of participants was 46.46 with a standard deviation of 16.33.
288 participants identified as male, while 308 identified as female, with 4 unknown.

\begin{table}
\caption{\label{tab:questtab}Likert Scale Questions asked of Study Participants}
\centering
\begin{tabu}{p{0.1\textwidth}p{0.8\textwidth}}
\toprule
Condition & Question\\
\midrule
All & How strong would you say the case against the defendant is?\\\\
All & How strong is the evidence that the defendant's gun was used to fire the shot in the convenience store, in your opinion?\\\\
All & How credible did you find the testimony of Terry Smith (the firearm examiner)?\\\\
Algorithm & How credible did you find the testimony of Adrian Jones (the algorithm expert)?\\\\
Algorithm & How reliable do you think the firearm evidence in this case is?\\\\
All & How reliable do you think the firearm examiner's subjective opinion of the bullet comparison evidence is, in this case?\\\\
Algorithm & How reliable do you think the firearm algorithm evidence is,  in this case?\\\\
All & How scientific do you think the firearm examiner's subjective opinion of the bullet comparison evidence is, in this case?\\\\
Algorithm & How scientific do you think the firearm evidence is in this case, overall?\\\\
Algorithm & How scientific do you think the firearm algorithm evidence is in this case?\\\\
All & Based on this testimony, how would you rate your understanding of the method described for the examiner's personal bullet comparison?\\\\
Algorithm & Based on this testimony, how would you rate your understanding of the method described for the bullet matching algorithm?\\\\
All & How often do firearm examiners make mistakes when determining whether bullets were fired through the same gun?\\\\
All & In general, how reliable do you think firearm evidence is?\\\\
All & In general, how scientific do you think firearm evidence is?\\
\bottomrule
\end{tabu}
\end{table}

\Cref{tab:questtab} summarizes the Likert scale questions asked of participants.
Participants were asked to rate the following according to a 7-point Likert scale: their views on the examiner's credibility as well as the evidence's reliability and scientificity (eg. ``extremely unreliable'' to ``extremely reliable'').
When the algorithm was absent, participants were asked to rate the reliability and scientificity of the examiner's firearm comparison and the field of firearm comparison as a whole.
When the algorithm was present, participants were additionally asked to judge the reliability of the algorithm comparison and the overall firearm comparison (including both the algorithm and the examiner's comparison) in addition to the examiner's firearm comparison and the field of firearm comparison as a whole.

Strength of evidence (e.g.~how much evidence there was to suggest the defendant was innocent or guilty) was measured on a 9-point Likert scale.

\hypertarget{scale-compression}{%
\subsection{Scale Compression}\label{scale-compression}}

\begin{figure}

{\centering \includegraphics[width=.9\textwidth]{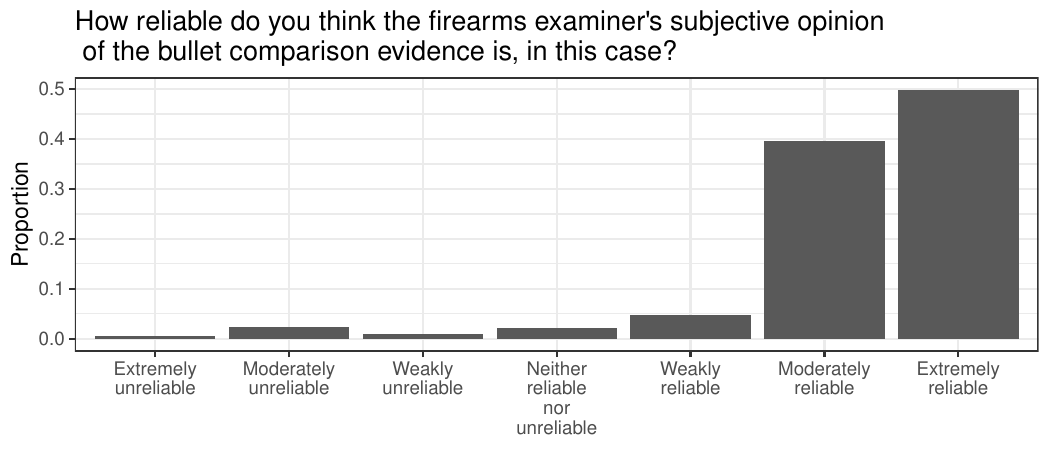} 

}

\caption{Histogram of perceived reliability of the exmaminer's firearm comparison across all study conditions. There is evidence of significant scale compression across conditions, suggesting that in order to be able to measure  and statistically model differences in perception of examiner reliability, the transcripts must contain more information which might cause participants to question the reliability of the examiner and of firearms comparisons.}\label{fig:reliablecompression}
\end{figure}

Throughout this analysis, likert-style responses commonly display \textbf{scale compression} - most of the participants' responses fall into one or two bins at one extreme of the scale.
This suggests that the scenario itself is not calibrated to be able to detect changes in participant views.
Scale compression may occur in quantitative responses if, for example, an exam designed to assess student learning results in a class average of 95: the test is not designed to be able to separate the students who understand the material extremely well from the students who have only partially mastered the material.
Our participants generally found that the examiner was credible and that the evidence presented was reliable and scientific.
This compression is demonstrated in \Cref{fig:reliablecompression}, which shows participant selection for reliability categories of the examiner's comparison across all experimental conditions.
Here, the vast majority of participants chose the top two categories (508 out of 569, or 89.28\%).
The other Likert categories have few observations - making it difficult to conduct comparisons across experimental conditions.
This lack of variation in responses makes it difficult to use standard statistical approaches, such as linear models.
In this paper, we use graphics to explore the data from this study in order to develop testable hypotheses for future iterations of studies using the Cole scenario.
In general, we primarily see strong effects when examining questions relating the examiner's conclusion (Identification/Match, Inconclusive, Elimination/Non match) to the likelihood that the gun was used in the crime or that the defendant was guilty.
This suggests that at the bare minimum, the scenario is well calibrated to assess the relationship between the examiner conclusion and the guilt or innocence of the suspect.

While we used a scenario which had been previously used in similar experiments \citep{garrett2018} and modified it to examine the use of algorithms, the original authors of this scenario are not statisticians and did not fully identify the underlying issues that may have contributed to an inability to detect differences between treatment groups in the original experiment.
It is important to design and calibrate these types of user experiments so that presented evidence is ``just right'' - not too strong, not too weak.
This calibration allows any manipulations, such as those in this study, to show up in the resulting data.
When scale compression is present, however, it is hard to show increased confidence in the examiner's opinion when the base assessment is already at ``extremely reliable''.
Thus, future iterations of this study need to do more to challenge the examiner's reliability and the perceived scientificity of the discipline - not because these things are necessarily in question (though there have been several successful legal challenges on the use of firearms evidence in court), but because in order to understand the effects of external manipulations, it is important to set up an experiment where these effects can be measured.
Inclusion of jury instructions (reminding the jury that they are the triers of fact and must be convinced beyond a reasonable doubt) and stronger cross-examination which focuses on error rates of firearms examination and examples of false convictions related to firearms may help to reduce scale compression and provide a more nuanced view of the effect of other interventions.

\hypertarget{credibility}{%
\subsection{Credibility}\label{credibility}}

\begin{figure}

{\centering \includegraphics[width=.9\textwidth]{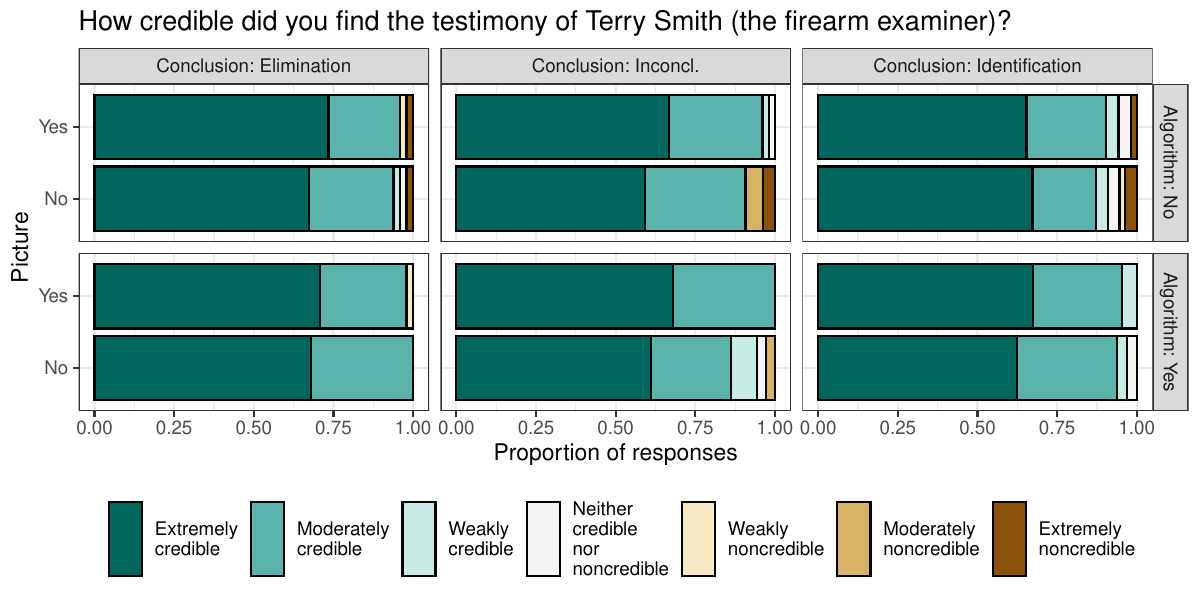} 

}

\caption{Histogram of examiner credibility by conclusion, demonstrative evidence, and algorithm conditions. Inconclusive conclusions have slightly lower credibility (particularly in the absence of demonstrative evidence), but overall, the primary observation when considering this data is that there is significant scale compression.}\label{fig:credible}
\end{figure}

All participants (regardless of experimental condition) were asked to rate the credibility of the firearms examiner; results are provided in \Cref{fig:credible}.
535 individuals selected ``moderately credible'' and ``extremely credible'' in the scale, while 34 individuals selecting lower categories.

In addition, participants who were assigned to the algorithm condition were asked to rate the credibility of the algorithm expert.
These results also show scale compression: 250 chose ``extremely credible'' or ``moderately credible'' and 9 chose a lower category.

\hypertarget{reliability}{%
\subsection{Reliability}\label{reliability}}

\begin{figure}

{\centering \includegraphics[width=\linewidth]{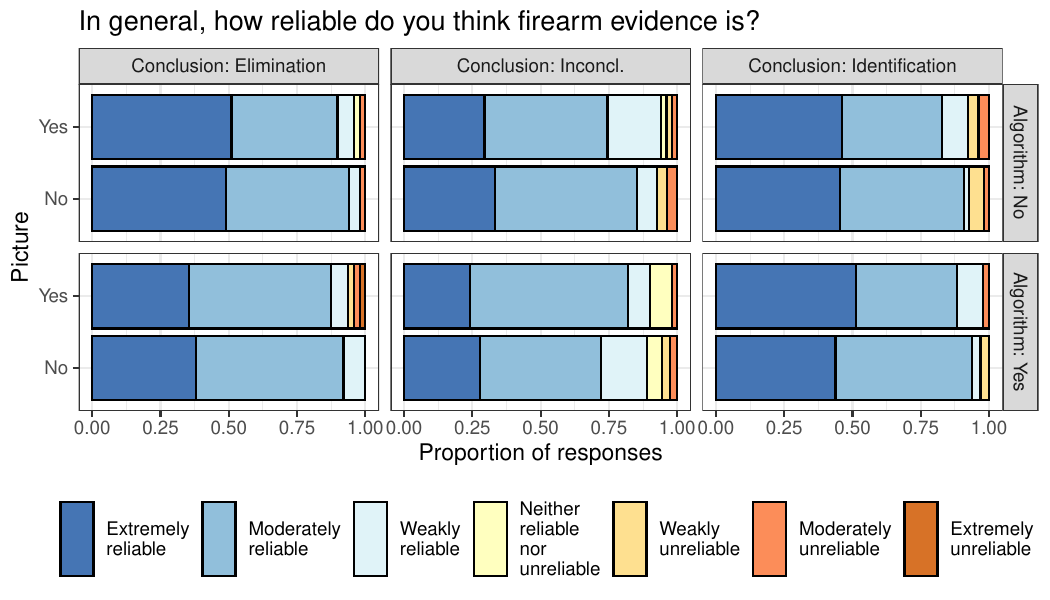} 

}

\caption{Perception of reliability of the field of firearms examination when manipulating demonstrative evidence (pictures) and use of the algorithm. Participants who read testimony about the algorithm were slightly more likely to say the field was moderately reliable and less likely to say the field was extremely reliable. Similarly, demonstrative evidence may be associated with a small reduction in perception of the reliability of the field.}\label{fig:reliable}
\end{figure}

\Cref{fig:reliable} displays the results for the perception of firearm reliability as a field; a question that was also asked of all participants.
This chart is similar to the graphic for credibility in that the top two categories (``moderately reliable'' and ``extremely reliable'') contain many of the responses (490 observations), while other lower categories are sparsely populated (79 observations).
In this case, however, there is some correlation between the conclusion and the ratings of reliability - specifically, both in the presence and absence of the algorithm, ``moderately reliable'' was more popular than ``extremely reliable'' for inconclusive decisions (for a cumulative total of 96 and 55 observations, respectively).
This is the opposite of the trend seen in the elimination and identification conditions, where ``moderately reliable'' contains similar or less observations than ``extremely reliable'' (93 and 85 respectively in the elimination condition; 76 and 85 respectively in the identification condition).

\hypertarget{scientificity}{%
\subsection{Scientificity}\label{scientificity}}

\begin{figure}

{\centering \includegraphics[width=\linewidth]{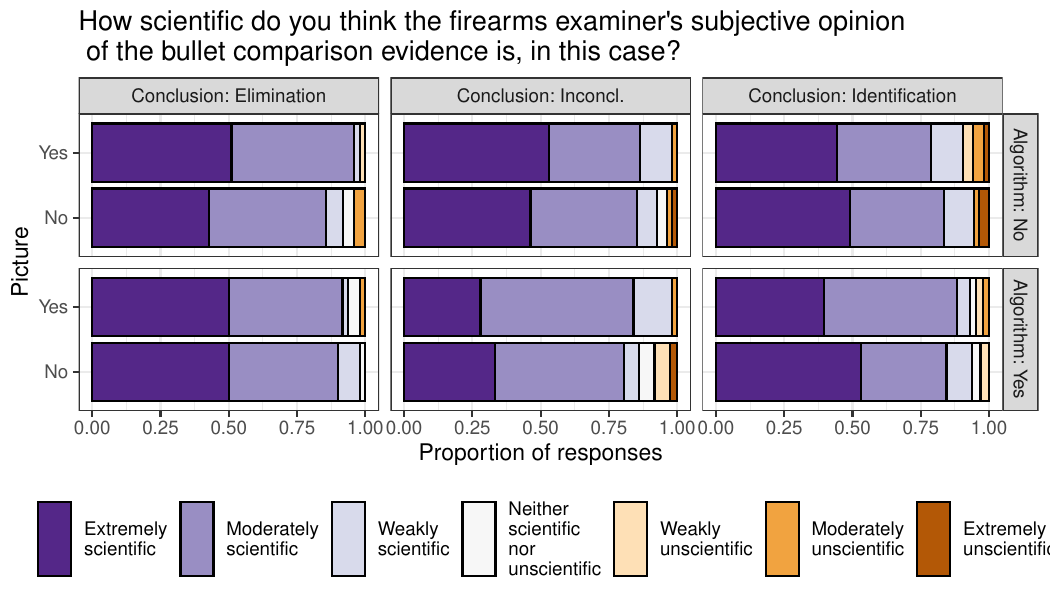} 

}

\caption{Perceived scientificity of the examiner's opinion by examiner conclusion, use of demonstrative evidence, and use of the algorithm. There are relatively few differences in perceived scientificity of the examiner's opinion across conditions, though inconclusive opinions seem to reduce perceptions of scientificity in the presence of the algorithm.}\label{fig:science}
\end{figure}

\Cref{fig:science} shows the opinion of the scientificity of the examiner's comparison, divided by algorithm condition.
As before, most responses are at the high end of the scale, with 491 observations in the two highest categories, and 78 observations in the remaining categories.
When evaluating the top two categories of scientificity, the inconclusive category demonstrates a different trend than the other categories when the algorithm is present.
While other categories either favor ``extremely scientific'' over ``moderately scientific'' or have approximately equal results, when the algorithm is present and there is an inconclusive decision, participants tended toward ``moderately scientific'' over ``extremely scientific'' (45 observations and 26 observations, respectively).

\hypertarget{understanding}{%
\subsection{Understanding}\label{understanding}}

\begin{figure}

{\centering \includegraphics[width=\textwidth]{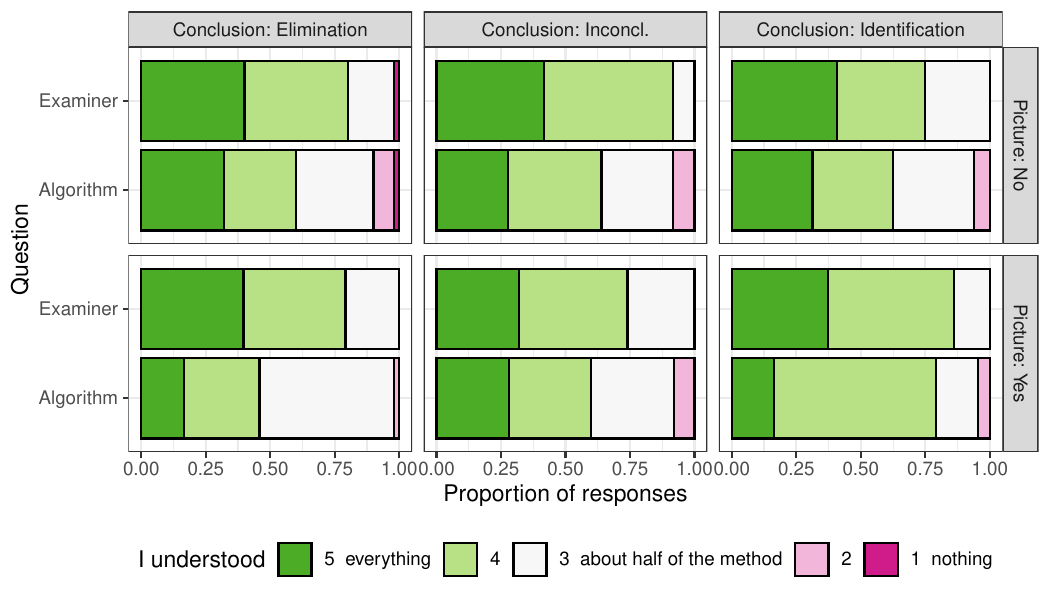} 

}

\caption{For the algorithm condition, participant understanding of algorithm and examiner testimony by conclusion and picture inclusion. Across conditions, participants were less confident in their understanding of the algorithm testimony compared to the examiner testimony.}\label{fig:histunder}
\end{figure}

\begin{table}

\caption{\label{tab:undertab}Table of participant understanding for participants in the algorithm condition, where 1 corresponds to understanding nothing and 5 corresponds to understanding everything.}
\centering
\begin{tabular}[t]{l|l|r|r|r|r|r}
\hline
Question & Conclusion & 1 & 2 & 3 & 4 & 5\\
\hline
Algorithm & Elimination & 1 & 5 & 40 & 28 & 24\\
\hline
Algorithm & Inconcl. & 0 & 7 & 26 & 29 & 24\\
\hline
Algorithm & Identification & 0 & 4 & 17 & 37 & 17\\
\hline
Examiner & Elimination & 1 & 0 & 19 & 39 & 39\\
\hline
Examiner & Inconcl. & 0 & 0 & 16 & 39 & 31\\
\hline
Examiner & Identification & 0 & 0 & 14 & 32 & 29\\
\hline
\end{tabular}
\end{table}

\begin{figure}

{\centering \includegraphics[width=\textwidth]{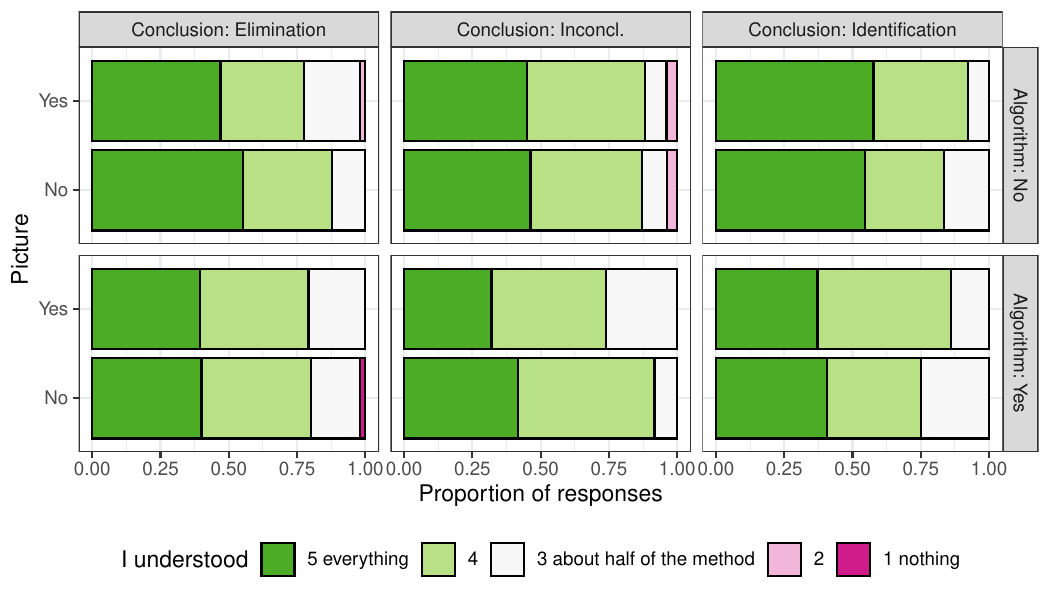} 

}

\caption{Histogram of perceived understanding of the firearms examiner's bullet comparison}\label{fig:underalgexp}
\end{figure}

Participants were asked to rate their understanding of the firearms examiner's comparison, as well as their understanding of the algorithm (when present).
These results are shown in \Cref{fig:histunder} and \Cref{tab:undertab} for those given the algorithm condition - comparing their rated understanding of the firearms examiner's comparison and the algorithm comparison when both are present.
Here, individuals mostly selected the three highest categories (242 vs 17 for the understanding of the algorithm and 258 vs 1 for the understanding of the forensics examiner's comparison), and there appears to be a difference in participants' rating of the algorithm and examiner when the algorithm is present.
The algorithm was generally assigned lower values of understanding than the examiner.
It should be noted that participants' rating of their own understanding may not truly indicate the understanding of the participants.
\citet{dunningFlawedSelfAssessmentImplications2004} found that an individual's rating of their own knowledge may not directly correspond to their actual learning level.
\Cref{fig:underalgexp} shows the perceived understanding of the firearms examiner's comparison across all categories.

\hypertarget{probability}{%
\subsection{Probability}\label{probability}}

\begin{figure}

{\centering \includegraphics[width=\textwidth]{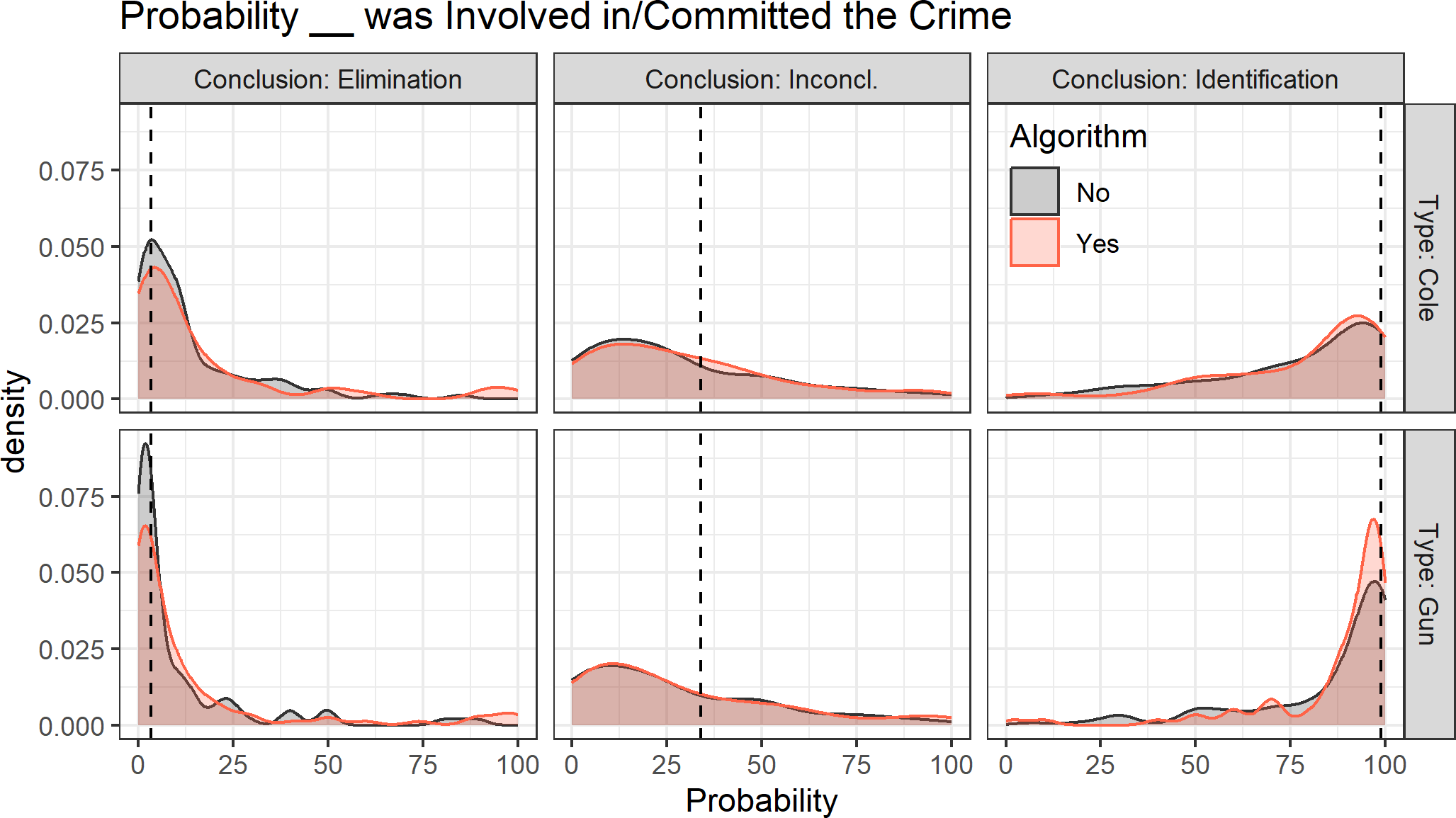} 

}

\caption{Probability the gun was used in the crime, or that Cole committed the crime. Dashed lines indicate bullet match scores for the algorithm. Participants were less committed to Cole's guilt or innocence than to the gun's involvement (or not) in the crime across all conditions, suggesting that at least some participants correctly discerned that the gun was not necessarily an indication of Cole's involvement, e.g. that the evidence linking him to the crime was circumstantial.}\label{fig:prob}
\end{figure}

Linear models are used for questions regarding the probability that Cole committed the crime and the probability that the gun was used in the crime.
These probabilities are modeled with a beta generalized linear model that considers the interaction between conditions; results can be found in the supplementary material.
\Cref{fig:prob} displays participant reported probabilities that Cole committed the crime or Cole's gun was involved in the crime.
This figure indicates that there is a difference between conclusions (as expected, due to differences in the strength of evidence presented), but not much of a difference for cases when the algorithm is included.
Here, in the case of an elimination, probabilities assigned by participants appear to be extremely low, while in the case of an identification, probabilities appear to be relatively high.
In the case of an inconclusive decision, participants also tended to favor a lower probability, although this trend is not as extreme as in the case of an elimination.
Because bullet match scores, resulting from the use of the algorithm, range from 0 to 1, there was concern that participants may incorrectly interpret this value as the probability of involvement in the crime.
The vertical lines in \Cref{fig:prob} indicate this match score for each condition.
There does not appear to be evidence that those who received the algorithm condition anchored to the match score values, when comparing the distribution of those who did not receive the match score (the non-algorithm condition).

\begin{figure}

{\centering \includegraphics[width=\textwidth]{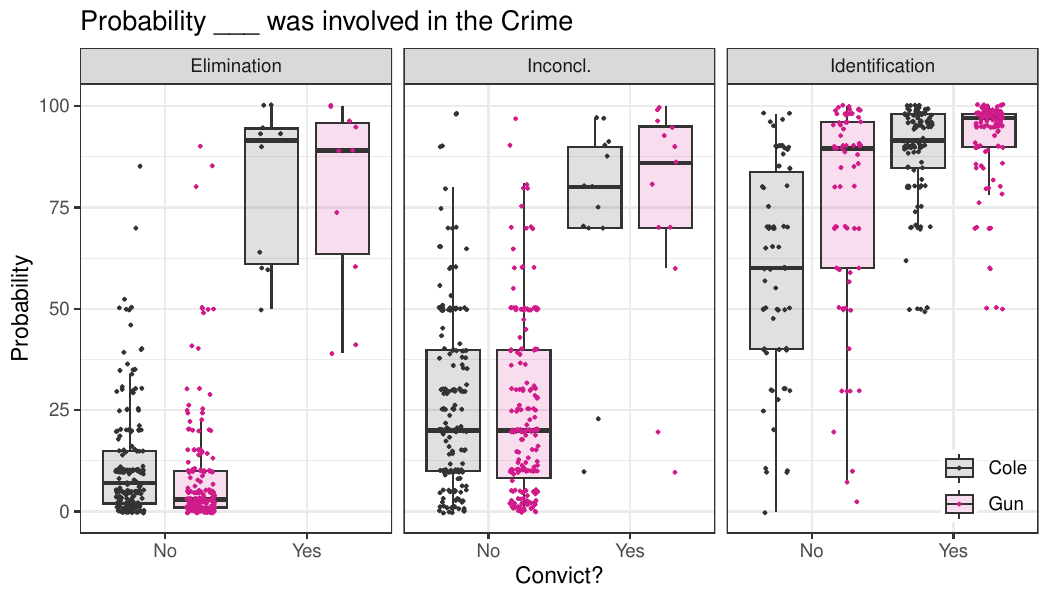} 

}

\caption{Probability that Cole committed the crime (black) and that the gun was used in the crime (pink) based on whether the participants chose to convict. Overall, participants responded to the testimony presented (Identification, Inconclusive, Elimination) in a reasonable manner. The largest discrepancy between probabilities for Cole and probabilities for the gun was in Identification; that is, many participants recognized that information about the gun's use (or not) in the crime may not provide the same level of information about Cole's guilt or innocence.}\label{fig:probguilt}
\end{figure}

Participants were also asked whether or not they would choose to convict Cole, based on the criminal trial standard of ``beyond a reasonable doubt''.
The conviction decision of participants and their assigned probabilities of involvement in the crime are shown in \Cref{fig:probguilt}, grouped by the conclusion of the firearms examiner.
10 out of 196 (5\%) who received the elimination condition, 13 out of 191 (7\%) who received the inconclusive decision, and 112 out of 182 (62\%) who received the identification condition chose to convict.
The few individuals who chose to convict in the elimination and inconclusive conditions assigned higher probabilities than their counterparts, indicating that individuals thought Cole was guilty or that the gun was used in the crime.
This discrepancy between conclusion and participant selections may be due to faulty internal logic, or a general belief that someone whose case has proceeded to trial is unlikely to be innocent, regardless of the evidence.
When the examiner made an identification, participants who chose to convict assigned higher probabilities that Cole committed the crime than those who did not convict.
The choice to not convict in the identification condition may be due to the ``beyond a reasonable doubt'' threshold: the only evidence that participants read matched the gun to the crime scene, but did not have evidence that tied Cole specifically to the crime.
This distinction between the gun and Cole can be seen in the relationship between those who chose not to convict when the bullets matched in \Cref{fig:probguilt}: they overall assigned a higher probability that gun was used in the crime than they did that Cole was involved in the crime (mean values of 75.57 and 60.06 with standard deviations of 25.94 and 25.47, respectively).

\hypertarget{summary}{%
\subsection{Summary}\label{summary}}

A common feature in many of these charts is scale compression - most individuals limited their Likert scale selection to the two highest values in terms of credibility, reliability, and scientificity.
This demonstrates that, across all experimental conditions, participants perceive the examiner as credible, and the evidence as reliable and scientific.
In this study, we were unable to discern a difference in perception of reliability, credibility, or scientificity between the algorithm and non-algorithm conditions or between demonstrative evidence and standard testimony conditions, though there are suggestions that effects may be present but not detectable due to scale compression (e.g.~\Cref{fig:underalgexp}).
Feelings regarding the strength of evidence, conviction decision, and probability Cole/the gun was involved in the crime varied by the conclusion of the firearms examiner, as expected.

The examiner's conclusion had the largest effect, both in the expected areas of strength of evidence and probability that Cole/the gun was involved in the crime, as well as in areas of reliability and scientificity.
There was also some difference between perceptions of the algorithm and the traditional bullet analysis method.
In particular, the explanation of the algorithm received lower scores of understanding than the explanation of the firearms examiner's bullet comparison.
This may be due to math aversion - terms used in the algorithm description sound difficult to understand, which may reduce the willingness of participants to try to parse the explanation, even if the explanation itself is not technical.

\hypertarget{discussion}{%
\section{Discussion}\label{discussion}}

In this study, questions using Likert scales were frequently subject to scale compression, complicating our attempts to statistically model responses by condition.
There are several solutions to this problem, but one of the most promising is to alter the scenario to introduce more doubt about the process of firearms examination, shifting participant answers towards the middle of the Likert scale.
Introducing discussion of erroneous convictions due to firearms evidence, adding more information about error rates in firearms examination, and including instructions to the jury from the judge before the examiner's testimony are all options which have occurred in real court cases and which would be expected to reduce scale compression and improve the statistical ability to discern effects of algorithms and demonstrative evidence on interpretation of firearms testimony.

Another solution is to move away from Likert scales to responses in different formats: probabilities, numerical chance, betting, and opinion of guilt.
We are in the process of executing a short follow-up study which examines different ways to ask these questions, and our goal is to explore not only how participant responses change, but to also demonstrate ways to model these different response types effectively.

Very few studies are executed perfectly; this study is no exception.
There were two minor mistakes in the transcript which were present for approximately the first half of participants.
These typos included referring to the firearms expert as Alex Smith as opposed to Terry Smith in all scenario questions, and for the cross examination in the elimination testimony.
In the case of non-algorithm inconclusive testimonies, the question: ``Can you describe the process of obtaining these test fired bullets?'' was missing, but the response: ``The test-fired bullets came from a test fire of the gun recovered from the traffic stop.'' remained unchanged.
There was no indication that participants were confused by these typos, but because Prolific recruits participants for separate demographic categories, these typos are confounded with the demographic variables, because blocks for demographics with higher participation on Prolific (younger ages, whites) fill up more quickly than blocks with lower participation on the site.

In addition to the data-driven modifications above, we also noticed during the execution of this study that participants may not have fully understood the difference between initial testimony and cross examination, or which witnesses were testifying for the prosecution vs.~the defense.
The transcript format provided in this study followed the same format as the court transcripts - speakers were indicated by ``Q:'' and ``A:'', but the identity of the speaker and their alignment within the courtroom could be easily confused.
To address this, we plan to implement a more visual representation of a courtroom transcript, using graphics to show each individual who is speaking and subtle cues to indicate which side they are testifying for.

We expect that these combined modifications will produce a study with more nuanced participant responses and will alleviate the scale compression seen in this experiment.

\hypertarget{acknowledgements}{%
\section{Acknowledgements}\label{acknowledgements}}

Thank you to the study participants, without whom this work would not be possible.

This work was funded (or partially funded) by the Center for Statistics and Applications in Forensic Evidence (CSAFE) through Cooperative Agreements 70NANB15H176 and 70NANB20H019 between NIST and Iowa State University, which includes activities carried out at Carnegie Mellon University, Duke University, University of California Irvine, University of Virginia, West Virginia University, University of Pennsylvania, Swarthmore College and University of Nebraska, Lincoln.

Computation and visualizations were made possible thanks to the following R packages: `tidyverse'\citep{tidyverse}, `RColorBrewer' \citep{colorbrewer}, `patchwork' \citep{patchwork}, `gt' \citep{gt}, `MASS' \citep{mass}, `emmeans' \citep{means}, `shiny' \citep{shiny}, and `mgcv' \citep{mgcv}.

\bibliography{bibliography.bib}
\bibliographystyle{jds}

\end{document}


\begin{frontmatter}

\title{Supplement to Demonstrative Evidence and the Use of Algorithms in Jury Trials}
\runtitle{Algorithms in Jury Trials}

\author[1]{
  \inits{R.}
  \fnms{Rachel}
  \snm{Rogers}  \thanksref{1}  \ead{rachel.rogers@huskers.unl.edu}}
\author[1]{
  \inits{S.}
  \fnms{Susan}
  \snm{VanderPlas}}

\thankstext[type=corresp,id=1]{Corresponding author}
\address[1]{Department of Statistics, 
  \institution{University of Nebraska-Lincoln}, \cny{United States of America}}

\begin{abstract}
We investigate how the use of bullet comparison algorithms and demonstrative evidence may affect juror perceptions of reliability, credibility, and understanding of expert witnesses and presented evidence. The use of statistical methods in forensic science is motivated by a lack of scientific validity and error rate issues present in many forensic analysis methods. We explore what our study says about how this type of forensic evidence is perceived in the courtroom -- where individuals unfamiliar with advanced statistical methods are asked to evaluate results in order to assess guilt. In the course of our initial study, we found that individuals overwhelmingly provided high Likert scale ratings in reliability, credibility, and scientificity regardless of experimental condition. This discovery of scale compression - where responses are limited to a few values on a larger scale, despite experimental manipulations - limits statistical modeling but provides opportunities for new experimental manipulations which may improve future studies in this area.
\end{abstract}

\begin{keywords}
\kwd{explainable machine learning}\kwd{jury perception}.
\end{keywords}

\end{frontmatter}

\hypertarget{ordinal-logistic-regression-likert-scales}{%
\section{Ordinal Logistic Regression (Likert Scales)}\label{ordinal-logistic-regression-likert-scales}}

Because there are not enough observations in all categories, only categories with enough observations are considered.
Most analyses are also limited to main effects.
Due to the scale compression mentioned throughout the article, this analytical approach is not recommended, as it ignores key aspects of the data collection process (such as the inclusion of the complete scale).
If there are only two categories for consideration, the response is considered as binomial.
If there are more than two categories, ordered logistic regression is first implemented using the `VGAM' package (the `polr' package implementation failed to find starting values in several cases), and assumptions of proportional odds are tested by comparing the likelihood to the model without the parallel odds assumption.
Unless otherwise noted, the parallel odds assumption holds.
In a few cases, there were not enough observations for the model to be computed without the parallel odds assumption.

\hypertarget{credibility}{%
\subsection{Credibility}\label{credibility}}

\hypertarget{how-credible-did-you-find-the-testimony-of-terry-smith-the-firearm-examiner}{%
\subsubsection{How credible did you find the testimony of Terry Smith (the firearm examiner)?}\label{how-credible-did-you-find-the-testimony-of-terry-smith-the-firearm-examiner}}

\begin{figure}

{\centering \includegraphics{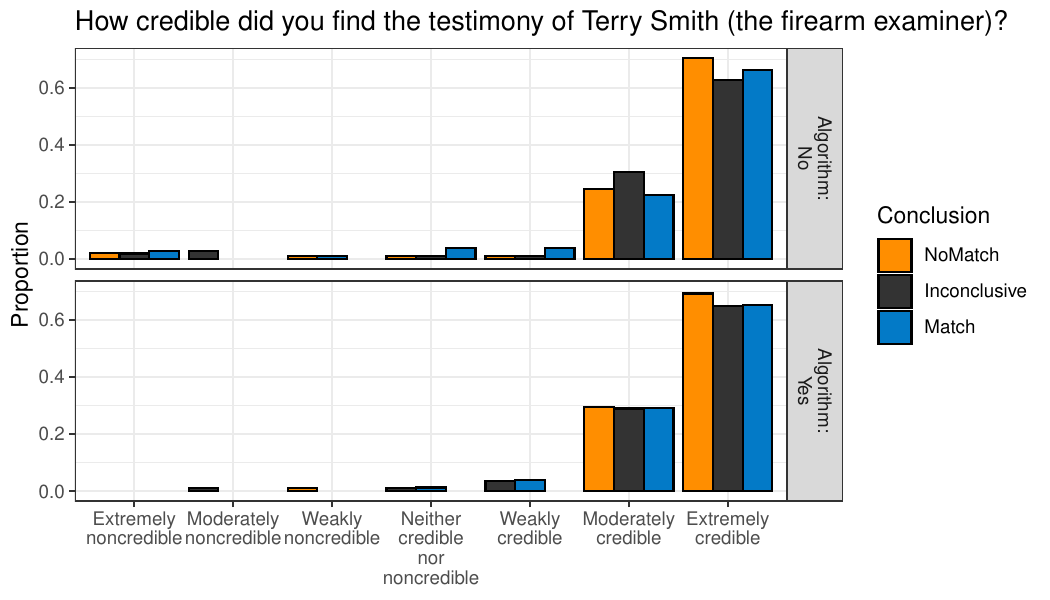} 

}

\caption{Histogram of Firearms Examiner Credibility}\label{fig:credible}
\end{figure}

Figure \ref{fig:credible} indicates that only the top two categories of the Likert scale have enough data for a formal analysis.
Thus, only the top two categories were considered using a binomial generalized linear model.
This model does not fully consider the responses or response options given to participants, and is not recommended.
There were not significant differences between conditions.

\begin{verbatim}
## Analysis of Deviance Table
## 
## Model: binomial, link: logit
## 
## Response: firetestcred
## 
## Terms added sequentially (first to last)
## 
## 
##            Df Deviance Resid. Df Resid. Dev Pr(>Chi)
## NULL                         534     645.82         
## Algorithm   1  0.41865       533     645.40   0.5176
## Conclusion  2  0.95523       531     644.44   0.6203
## Picture     1  0.19970       530     644.24   0.6550
\end{verbatim}

\hypertarget{how-credible-did-you-find-the-testimony-of-adrian-jones-the-algorithm-expert}{%
\subsubsection{How credible did you find the testimony of Adrian Jones (the algorithm expert)?}\label{how-credible-did-you-find-the-testimony-of-adrian-jones-the-algorithm-expert}}

\begin{figure}

{\centering \includegraphics{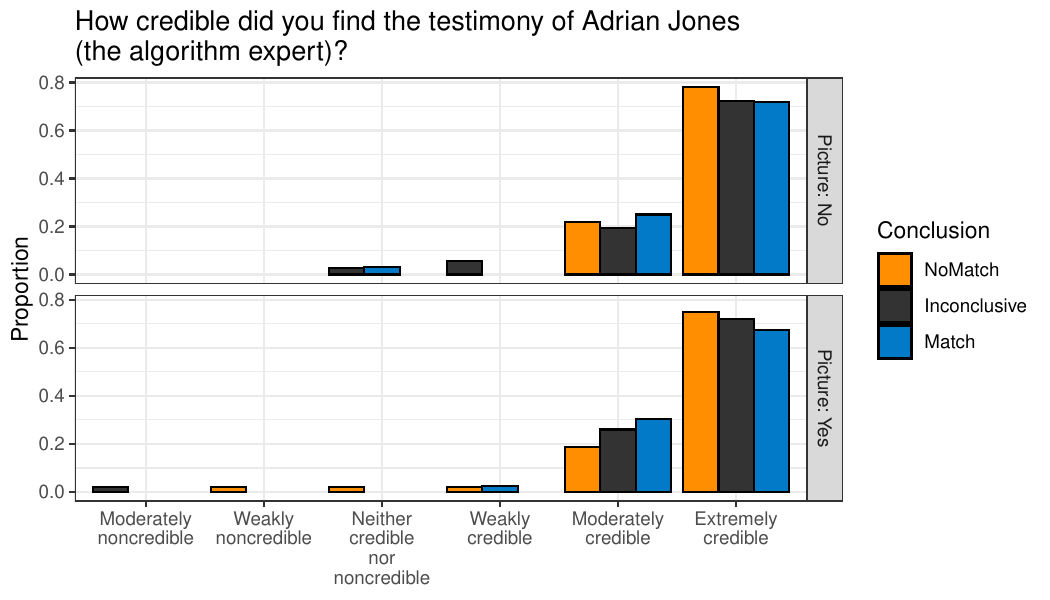} 

}

\caption{Histogram of Algorithm Expert Credibility}\label{fig:algcred}
\end{figure}

Similar to Figure \ref{fig:credible}, Figure \ref{fig:algcred} shows that most individuals only selected the top two categories of the Likert scale.
Thus, as before, only the top two categories will be considered in statistical analysis (although this does not reflect how the data was collected).

\begin{verbatim}
## Analysis of Deviance Table
## 
## Model: binomial, link: logit
## 
## Response: algtestcred
## 
## Terms added sequentially (first to last)
## 
## 
##            Df Deviance Resid. Df Resid. Dev Pr(>Chi)
## NULL                         249     277.82         
## Conclusion  2  1.32358       247     276.50   0.5159
## Picture     1  0.19895       246     276.30   0.6556
\end{verbatim}

\hypertarget{reliability}{%
\subsection{Reliability}\label{reliability}}

\hypertarget{in-general-how-reliable-do-you-think-firearm-evidence-is}{%
\subsubsection{In general, how reliable do you think firearm evidence is?}\label{in-general-how-reliable-do-you-think-firearm-evidence-is}}

\begin{figure}

{\centering \includegraphics{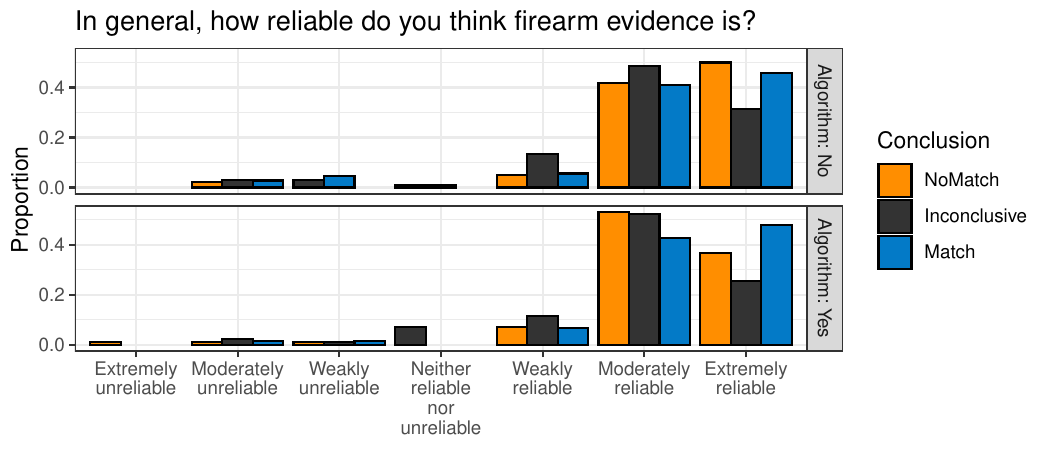} 

}

\caption{Histogram of perceived firearm reliability as a field}\label{fig:reliable}
\end{figure}

Figure \ref{fig:reliable} has observations from each condition combination in the top three categories of the Likert scale (weakly reliable, moderately reliable, and extremely reliable), so an ordered logistic regression using the three top categories is used.

\begin{verbatim}
##                           Estimate Std. Error    z value     Pr(>|z|)
## (Intercept):1           2.69064642  0.2301020 11.6932783 1.379566e-31
## (Intercept):2          -0.04658948  0.1823326 -0.2555192 7.983221e-01
## ConclusionInconclusive -0.63775184  0.2057890 -3.0990566 1.941379e-03
## ConclusionMatch         0.14463400  0.2051155  0.7051343 4.807267e-01
## PictureYes             -0.04996001  0.1683530 -0.2967574 7.666517e-01
## AlgorithmYes           -0.23425313  0.1693210 -1.3834855 1.665161e-01
\end{verbatim}

\hypertarget{how-reliable-do-you-think-the-firearm-evidence-in-this-case-is}{%
\subsubsection{How reliable do you think the firearm evidence in this case is?}\label{how-reliable-do-you-think-the-firearm-evidence-in-this-case-is}}

\begin{figure}

{\centering \includegraphics{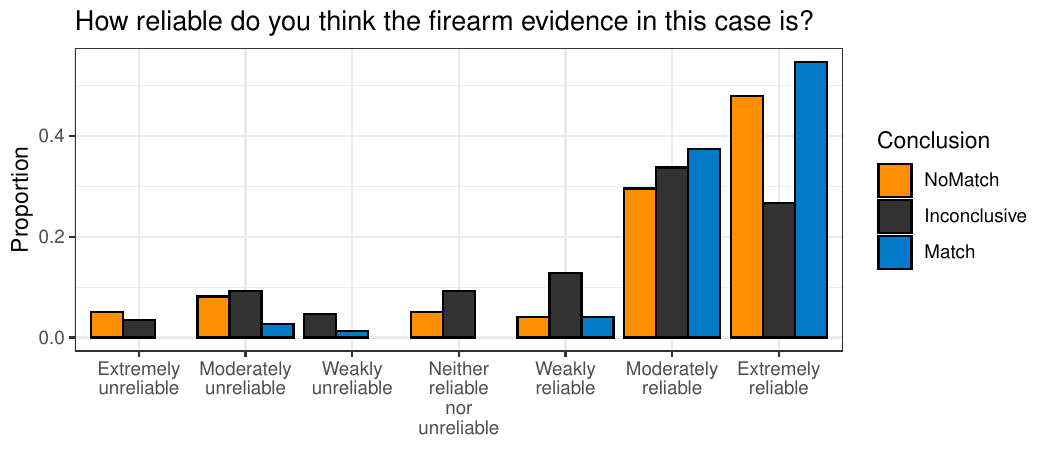} 

}

\caption{Histogram of overall case reliability}\label{fig:caserel}
\end{figure}

Based on Figure \ref{fig:caserel}, the top three categories contain results and are used in analysis (using ordered logistic regression).

\begin{verbatim}
##                           Estimate Std. Error    z value     Pr(>|z|)
## (Intercept):1           2.75992073  0.3442074  8.0181921 1.073129e-15
## (Intercept):2           0.33859995  0.2606274  1.2991725 1.938847e-01
## ConclusionInconclusive -1.02844224  0.3351858 -3.0682750 2.152984e-03
## ConclusionMatch        -0.05880294  0.3246099 -0.1811495 8.562502e-01
## PictureYes              0.06098488  0.2724282  0.2238567 8.228688e-01
\end{verbatim}

\hypertarget{how-reliable-do-you-think-the-firearms-examiners-subjective-opinion-of-the-bullet-comparison-is-in-this-case}{%
\subsubsection{How reliable do you think the firearms examiner's subjective opinion of the bullet comparison is, in this case?}\label{how-reliable-do-you-think-the-firearms-examiners-subjective-opinion-of-the-bullet-comparison-is-in-this-case}}

\begin{figure}

{\centering \includegraphics{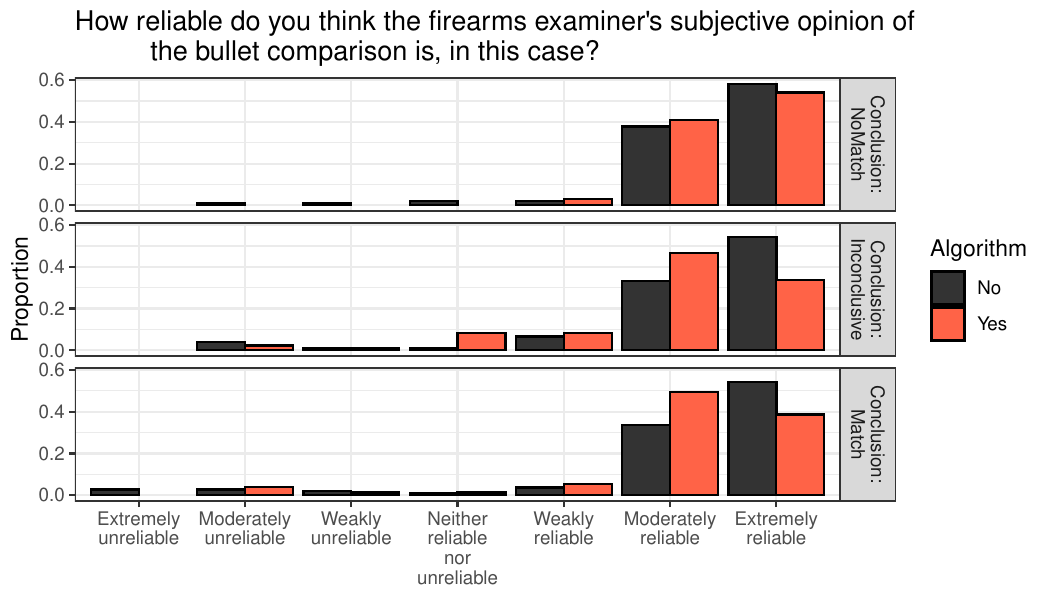} 

}

\caption{Histogram of perceived firearm exam reliability}\label{fig:examrel}
\end{figure}

In this case, there may not be enough observations in the ``weakly reliable'' category for analysis - so only the highest two categories of the Likert scale are analyzed.

\begin{verbatim}
##                           Estimate Std. Error    z value     Pr(>|z|)
## (Intercept):1           3.48591557  0.2692876 12.9449556 2.509082e-38
## (Intercept):2           0.62481123  0.1899804  3.2888202 1.006083e-03
## ConclusionInconclusive -0.43256548  0.2083372 -2.0762755 3.786847e-02
## ConclusionMatch        -0.28182315  0.2112147 -1.3342969 1.821066e-01
## PictureYes             -0.08513639  0.1718398 -0.4954406 6.202891e-01
## AlgorithmYes           -0.51673224  0.1729951 -2.9869753 2.817525e-03
\end{verbatim}

\hypertarget{how-reliable-do-you-think-the-firearm-algorithm-evidence-is-in-this-case}{%
\subsubsection{How reliable do you think the firearm algorithm evidence is, in this case?}\label{how-reliable-do-you-think-the-firearm-algorithm-evidence-is-in-this-case}}

\begin{figure}

{\centering \includegraphics{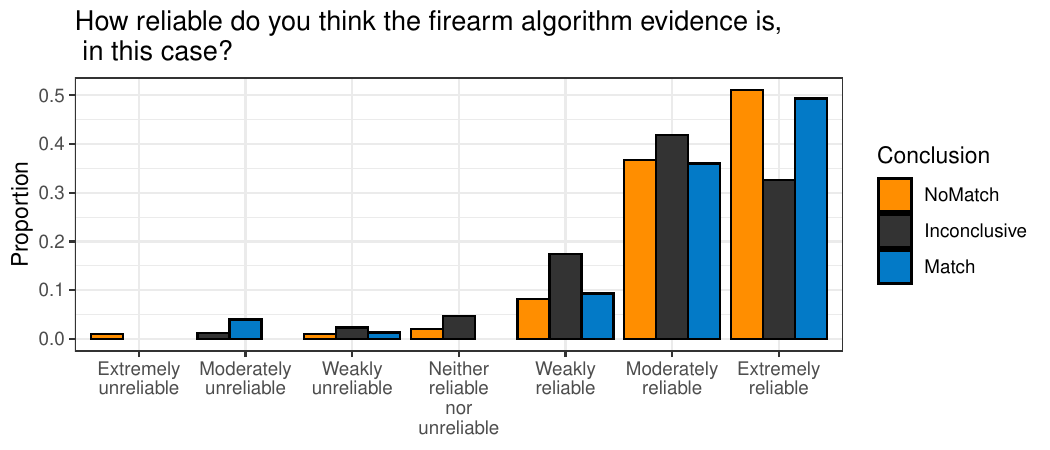} 

}

\caption{Histogram of perceived algorithm reliability}\label{fig:algrel}
\end{figure}

In Figure \ref{fig:algrel}, there are enough observations in the three highest categories for ordered logistic regression.

\begin{verbatim}
##                           Estimate Std. Error    z value     Pr(>|z|)
## (Intercept):1           2.53041886  0.3294040  7.6818090 1.568576e-14
## (Intercept):2           0.20609156  0.2570606  0.8017236 4.227129e-01
## ConclusionInconclusive -0.60818214  0.3317086 -1.8334830 6.673077e-02
## ConclusionMatch        -0.04176306  0.3195061 -0.1307113 8.960037e-01
## PictureYes             -0.10797218  0.2700328 -0.3998483 6.892682e-01
\end{verbatim}

\hypertarget{scientificity}{%
\subsection{Scientificity}\label{scientificity}}

\hypertarget{in-general-how-scientific-do-you-think-firearm-evidence-is}{%
\subsubsection{In general, how scientific do you think firearm evidence is?}\label{in-general-how-scientific-do-you-think-firearm-evidence-is}}

\begin{figure}

{\centering \includegraphics{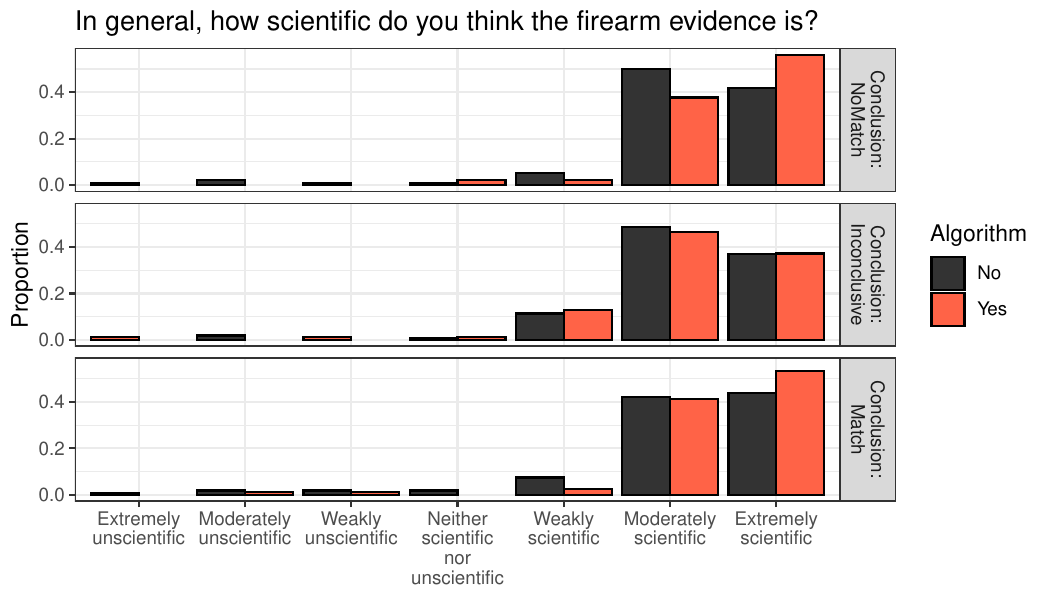} 

}

\caption{Histogram of perceived firearm scientificity as a field}\label{fig:gensci}
\end{figure}

The top three categories are used for analysis (Figure \ref{fig:gensci}).

\begin{verbatim}
##                           Estimate Std. Error    z value     Pr(>|z|)
## (Intercept):1           2.63271547  0.2326023 11.3185274 1.062415e-29
## (Intercept):2          -0.09899810  0.1834667 -0.5395972 5.894748e-01
## ConclusionInconclusive -0.60334328  0.2027738 -2.9754504 2.925587e-03
## ConclusionMatch        -0.03547145  0.2063989 -0.1718587 8.635486e-01
## PictureYes              0.02853508  0.1672198  0.1706442 8.645036e-01
## AlgorithmYes            0.32016403  0.1683871  1.9013569 5.725528e-02
\end{verbatim}

\hypertarget{how-scientific-do-you-think-the-firearm-evidence-in-this-case-is}{%
\subsubsection{How scientific do you think the firearm evidence in this case is?}\label{how-scientific-do-you-think-the-firearm-evidence-in-this-case-is}}

\begin{figure}

{\centering \includegraphics{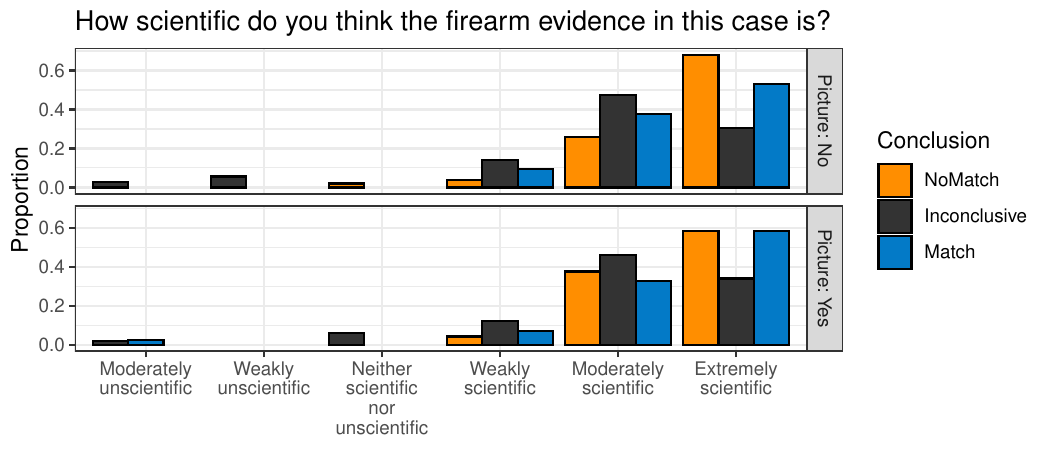} 

}

\caption{Histogram of perceived overall scientificity in this case}\label{fig:oversci}
\end{figure}

Shown in Figure \ref{fig:oversci}, the top three categories are considered for analysis.

\begin{verbatim}
##                          Estimate Std. Error    z value     Pr(>|z|)
## (Intercept)             2.6563989  0.5543173  4.7921995 1.649627e-06
## ConclusionInconclusive -1.8349916  0.6316622 -2.9050206 3.672289e-03
## ConclusionMatch        -0.8206576  0.6791730 -1.2083188 2.269247e-01
## PictureYes              0.1948171  0.4913303  0.3965096 6.917292e-01
\end{verbatim}

\hypertarget{how-scientific-do-you-think-the-firearms-examiners-subjective-opinion-of-the-bullet-comparison-is-in-this-case}{%
\subsubsection{How scientific do you think the firearms examiner's subjective opinion of the bullet comparison is, in this case?}\label{how-scientific-do-you-think-the-firearms-examiners-subjective-opinion-of-the-bullet-comparison-is-in-this-case}}

\begin{figure}

{\centering \includegraphics{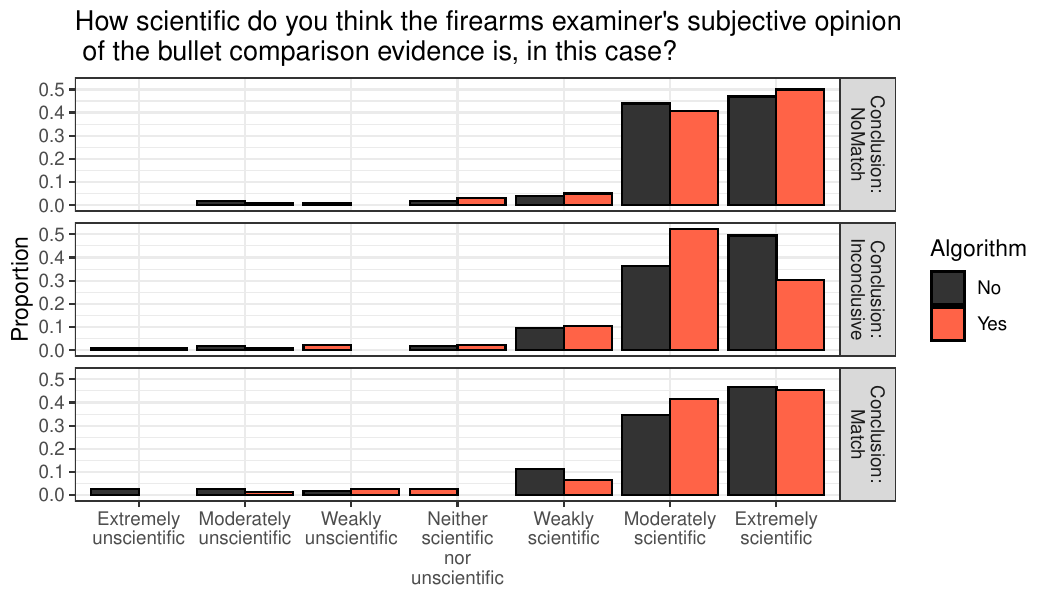} 

}

\caption{Histogram of perceived scientificity of the bullet comparison of the firearm examiner}\label{fig:science}
\end{figure}

Figure \ref{fig:science} shows this graph of scientificity.
There are observations in all three of the top categories, so an ordered logistic regression was used in this case as well.

\begin{verbatim}
##                            Estimate Std. Error     z value     Pr(>|z|)
## (Intercept)             2.405621778  0.4221794  5.69810255 1.211482e-08
## ConclusionInconclusive -0.961413670  0.4367253 -2.20141513 2.770665e-02
## ConclusionMatch        -0.772645299  0.4410679 -1.75176046 7.981501e-02
## PictureYes              0.009540393  0.3271550  0.02916169 9.767356e-01
## AlgorithmYes           -0.101246442  0.3330589 -0.30398963 7.611358e-01
\end{verbatim}

\hypertarget{how-scientific-do-you-think-the-firearm-algorithm-evidence-is-in-this-case}{%
\subsubsection{How scientific do you think the firearm algorithm evidence is, in this case?}\label{how-scientific-do-you-think-the-firearm-algorithm-evidence-is-in-this-case}}

\begin{figure}

{\centering \includegraphics{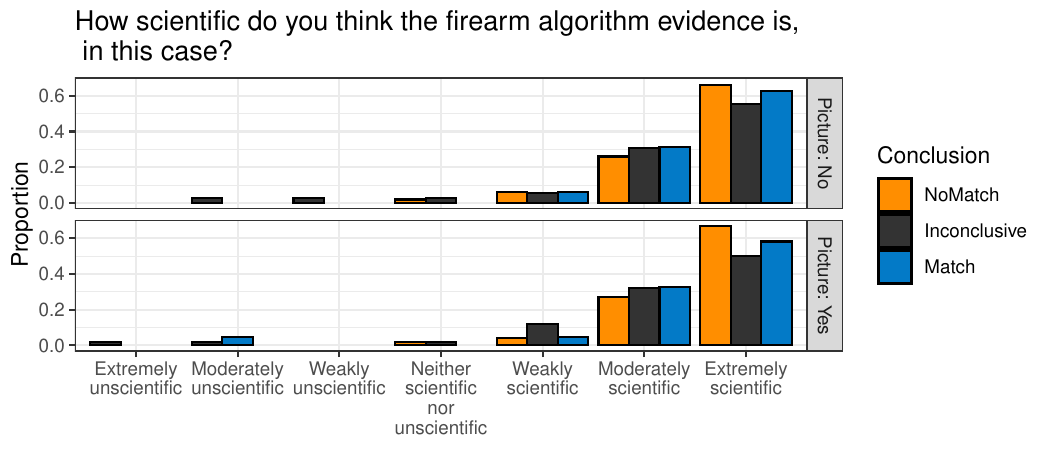} 

}

\caption{Histogram of perceived algorithm scientificity in this case}\label{fig:algsci}
\end{figure}

Figure \ref{fig:algsci} indicates that the top three categories may be used.

\begin{verbatim}
##                          Estimate Std. Error    z value     Pr(>|z|)
## (Intercept)             2.6593151  0.5396457  4.9278912 8.312189e-07
## ConclusionInconclusive -0.8200085  0.6042670 -1.3570299 1.747717e-01
## ConclusionMatch        -0.1325378  0.6992955 -0.1895304 8.496771e-01
## PictureYes             -0.1866046  0.5234277 -0.3565050 7.214624e-01
\end{verbatim}

\hypertarget{understanding}{%
\subsection{Understanding}\label{understanding}}

\hypertarget{based-on-this-testimony-how-would-you-rate-your-understanding-of-the-method-described-for-the-examiners-personal-bullet-comparison}{%
\subsubsection{Based on this testimony, how would you rate your understanding of the method described for the examiner's personal bullet comparison?}\label{based-on-this-testimony-how-would-you-rate-your-understanding-of-the-method-described-for-the-examiners-personal-bullet-comparison}}

\begin{figure}

{\centering \includegraphics{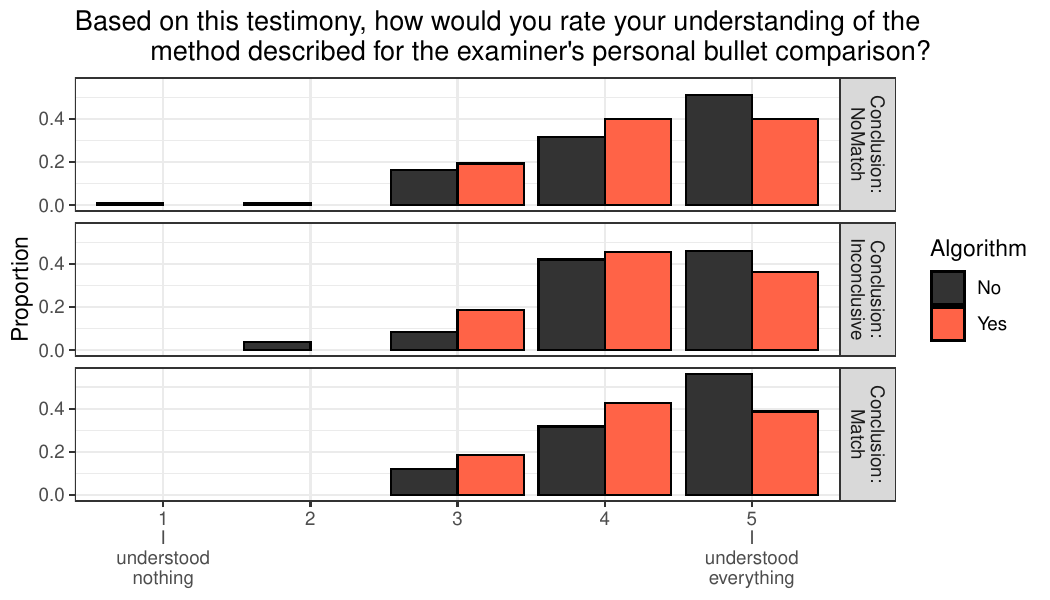} 

}

\caption{Histogram of understanding for the explanation of the firearms examiner}\label{fig:expunder}
\end{figure}

The top three categories are used, based on Figure \ref{fig:expunder}.

\begin{verbatim}
##                           Estimate Std. Error    z value     Pr(>|z|)
## (Intercept):1           2.00624872  0.1997633 10.0431280 9.850013e-24
## (Intercept):2           0.10125841  0.1770278  0.5719916 5.673277e-01
## ConclusionInconclusive -0.03983138  0.1939499 -0.2053695 8.372835e-01
## ConclusionMatch         0.11820234  0.1967775  0.6006902 5.480463e-01
## PictureYes             -0.12340507  0.1603832 -0.7694390 4.416328e-01
## AlgorithmYes           -0.52192165  0.1615615 -3.2304826 1.235814e-03
\end{verbatim}

\hypertarget{based-on-this-testimony-how-would-you-rate-your-understanding-of-the-method-described-for-the-bullet-matching-algorithm}{%
\subsubsection{Based on this testimony, how would you rate your understanding of the method described for the bullet matching algorithm?}\label{based-on-this-testimony-how-would-you-rate-your-understanding-of-the-method-described-for-the-bullet-matching-algorithm}}

\begin{figure}

{\centering \includegraphics{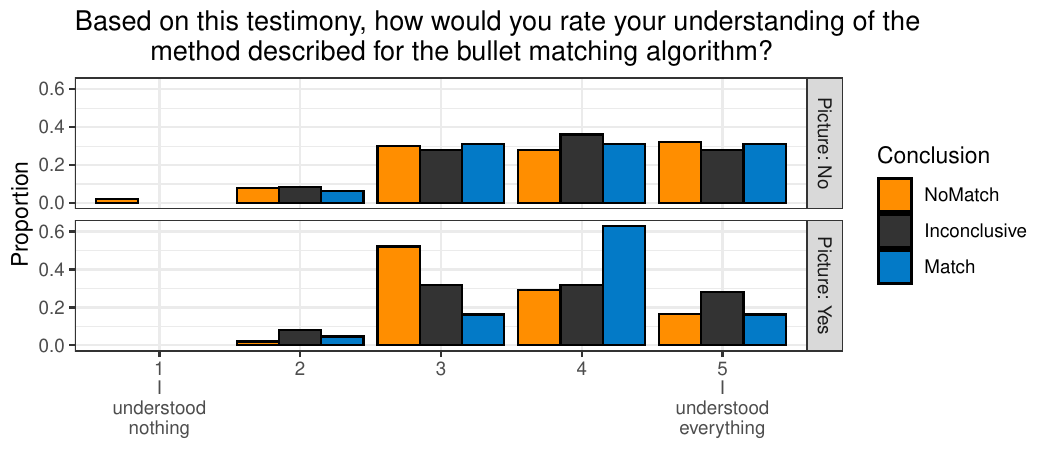} 

}

\caption{Histogram of understanding for the algorithm explanation}\label{fig:algunder}
\end{figure}

Here, all but the lowest category of understanding have observations, shown in Figure \ref{fig:algunder}.

\begin{verbatim}
##                          Estimate Std. Error    z value     Pr(>|z|)
## (Intercept):1           2.6771725  0.3169749  8.4460091 3.014302e-17
## (Intercept):2           0.4191797  0.2248766  1.8640433 6.231563e-02
## (Intercept):3          -1.1599859  0.2357655 -4.9200835 8.650731e-07
## ConclusionInconclusive  0.2285620  0.2714496  0.8420055 3.997849e-01
## ConclusionMatch         0.3907155  0.2822361  1.3843568 1.662492e-01
## PictureYes             -0.2358587  0.2294883 -1.0277590 3.040632e-01
\end{verbatim}

\hypertarget{strength}{%
\subsection{Strength}\label{strength}}

\hypertarget{strength-of-evidence-against-cole}{%
\subsubsection{Strength of Evidence against Cole}\label{strength-of-evidence-against-cole}}

\begin{figure}

{\centering \includegraphics{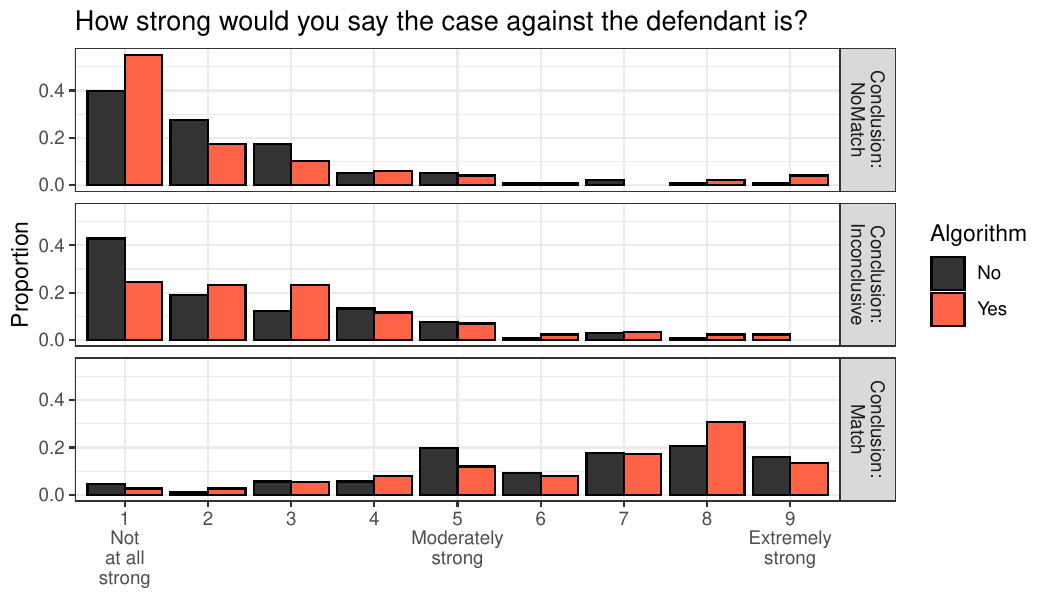} 

}

\caption{Histogram of perceived strength of evidence against the defendant}\label{fig:strength}
\end{figure}

Figure \ref{fig:strength} shows the participants' percieved strength of evidence against the defendant.
All categories are considered in this case.

\begin{verbatim}
##                          Estimate Std. Error     z value     Pr(>|z|)
## (Intercept):1           0.1736004  0.1720523   1.0089980 3.129756e-01
## (Intercept):2          -0.6988997  0.1756604  -3.9786986 6.929351e-05
## (Intercept):3          -1.4768982  0.1872769  -7.8861717 3.115979e-15
## (Intercept):4          -2.1048885  0.2016071 -10.4405452 1.618784e-25
## (Intercept):5          -2.9139033  0.2231067 -13.0605816 5.530366e-39
## (Intercept):6          -3.2467080  0.2314644 -14.0268143 1.068443e-44
## (Intercept):7          -3.9233876  0.2480721 -15.8155133 2.432194e-56
## (Intercept):8          -5.1296546  0.2882367 -17.7966758 7.499004e-71
## ConclusionInconclusive  0.5794179  0.1864103   3.1082934 1.881712e-03
## ConclusionMatch         3.6254826  0.2369178  15.3027046 7.334425e-53
## PictureYes             -0.3410194  0.1515947  -2.2495466 2.447774e-02
## AlgorithmYes            0.1443232  0.1521668   0.9484538 3.428985e-01
\end{verbatim}

\hypertarget{strength-of-evidence-against-the-gun}{%
\subsubsection{Strength of Evidence Against the Gun}\label{strength-of-evidence-against-the-gun}}

\begin{figure}

{\centering \includegraphics{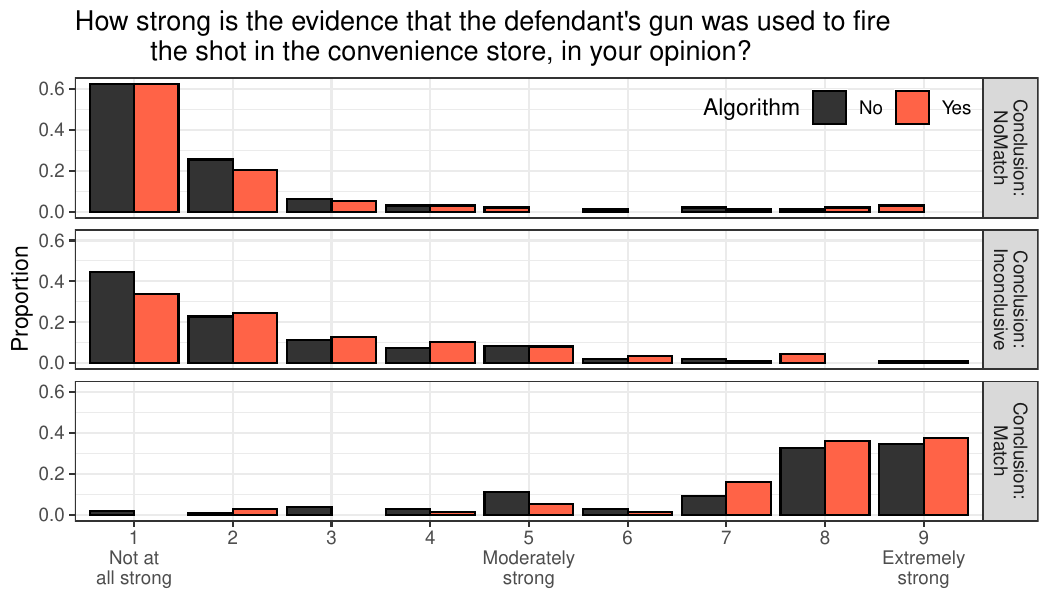} 

}

\caption{Histogram of perceived strength of evidence against the gun}\label{fig:gunstrength}
\end{figure}

In this case, there are not enough observations for the algorithm condition in each level for non-parallel odds (Figure \ref{fig:gunstrength}).
Therefore, only parallel odds are computed.
All levels are included in the model.

\begin{verbatim}
##                          Estimate Std. Error    z value     Pr(>|z|)
## (Intercept):1          -0.6243571  0.1819157  -3.432123 5.988753e-04
## (Intercept):2          -1.7057502  0.1988435  -8.578354 9.624035e-18
## (Intercept):3          -2.2727735  0.2128996 -10.675328 1.327877e-26
## (Intercept):4          -2.7732129  0.2287528 -12.123186 7.960358e-34
## (Intercept):5          -3.5055672  0.2562690 -13.679247 1.350838e-42
## (Intercept):6          -3.7292055  0.2642989 -14.109801 3.305072e-45
## (Intercept):7          -4.3357854  0.2837070 -15.282619 9.984606e-53
## (Intercept):8          -5.7081607  0.3157797 -18.076403 4.889575e-73
## ConclusionInconclusive  1.0040458  0.1972809   5.089423 3.591552e-07
## ConclusionMatch         5.0772852  0.2918746  17.395433 8.934839e-68
## PictureYes             -0.1014063  0.1574368  -0.644108 5.195054e-01
## AlgorithmYes            0.2987159  0.1587286   1.881929 5.984566e-02
\end{verbatim}

\hypertarget{mistakes}{%
\subsection{Mistakes}\label{mistakes}}

\begin{figure}

{\centering \includegraphics{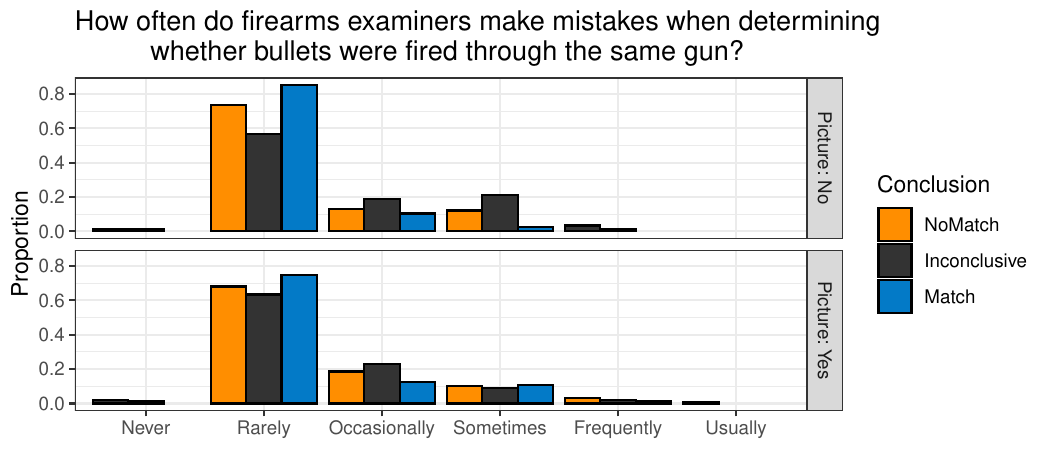} 

}

\caption{Histogram of perceived frequency of mistakes made by firearms examiners}\label{fig:mistakes}
\end{figure}

Figure \ref{fig:mistakes} shows the perceived frequency that firearms examiners make mistakes.
As can be seen, most individuals selected ``Rarely'', with little variations between conditions.
Very few people selected the extreme values of the scale - ``Never'' and ``Usually''.
Only the values of ``Rarely'', ``Occasionally'', and ``Sometimes'' are considered in the analysis.

\begin{verbatim}
##                           Estimate Std. Error      z value     Pr(>|z|)
## (Intercept):1          -1.30815117  0.2141733  -6.10790967 1.009445e-09
## (Intercept):2          -2.46653570  0.2386186 -10.33673008 4.806592e-25
## ConclusionInconclusive  0.47767483  0.2209376   2.16203507 3.061547e-02
## ConclusionMatch        -0.46577844  0.2516942  -1.85057254 6.423107e-02
## PictureYes              0.01634266  0.1914690   0.08535408 9.319799e-01
## AlgorithmYes            0.62735090  0.1924081   3.26052189 1.112074e-03
\end{verbatim}

\hypertarget{beta-distributions}{%
\section{Beta Distributions}\label{beta-distributions}}

\hypertarget{probability}{%
\subsection{Probability}\label{probability}}

\begin{figure}

{\centering \includegraphics[width=400px]{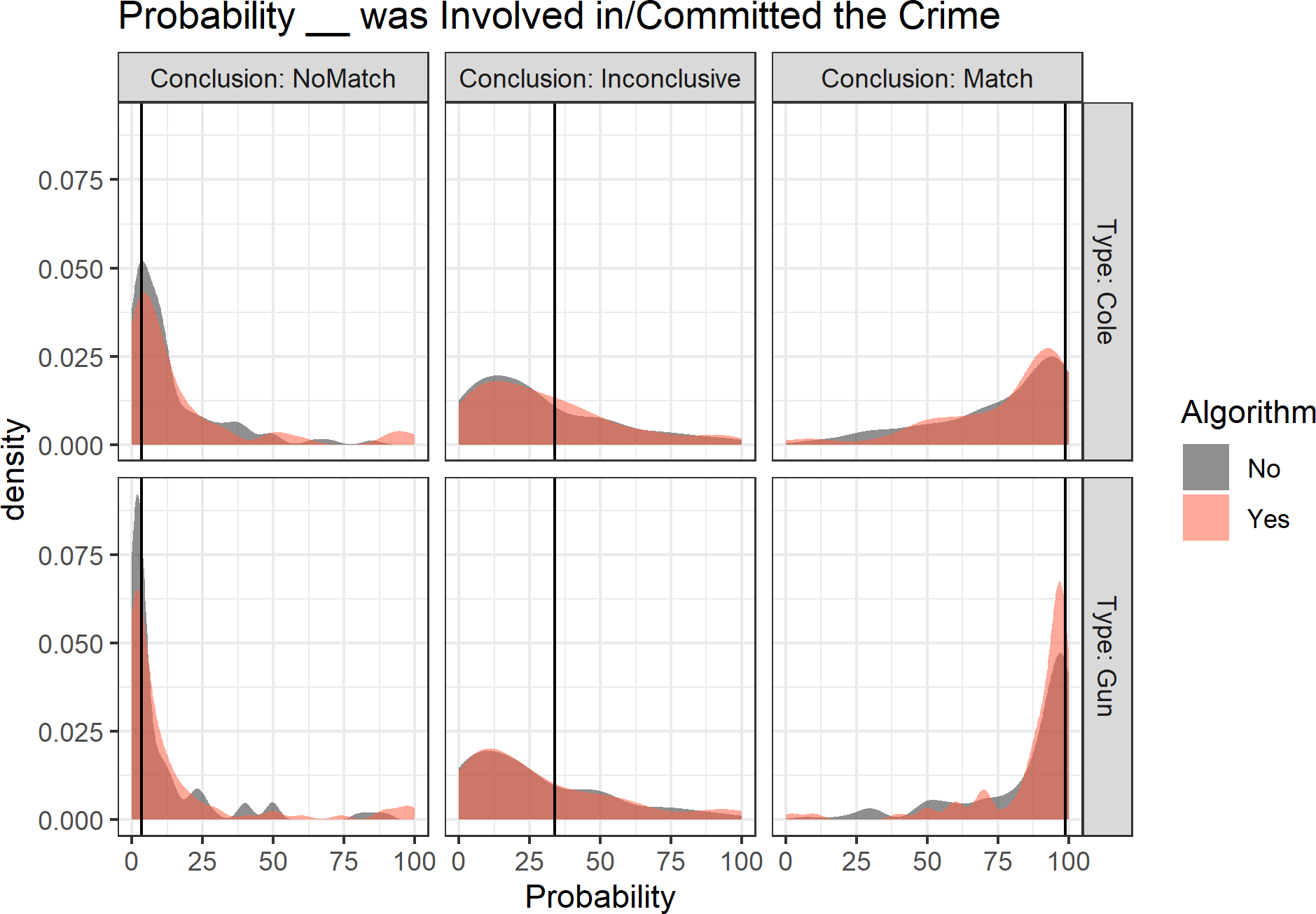} 

}

\caption{Probability the gun was used in the crime, or that Cole committed the crime. Black lines indicate bullet match scores for the algorithm.}\label{fig:prob}
\end{figure}

The graphs of probability are shown in Figure \ref{fig:prob}.
In both cases, the only significant effect is that of conclusion.
This analysis uses the `gam' function from the `mgcv' package.

\hypertarget{probability-cole-committed-the-crime}{%
\subsubsection{Probability Cole Committed the Crime}\label{probability-cole-committed-the-crime}}

\begin{verbatim}
##  model term                   df1 df2 F.ratio p.value
##  Conclusion                     2 557 263.793  <.0001
##  Picture                        1 557   0.209  0.6481
##  Algorithm                      1 557   0.007  0.9327
##  Conclusion:Picture             2 557   1.518  0.2201
##  Conclusion:Algorithm           2 557   0.422  0.6557
##  Picture:Algorithm              1 557   0.585  0.4446
##  Conclusion:Picture:Algorithm   2 557   0.149  0.8620
\end{verbatim}

\hypertarget{probability-the-gun-was-involved-in-the-crime}{%
\subsubsection{Probability the Gun was Involved in the Crime}\label{probability-the-gun-was-involved-in-the-crime}}

\begin{verbatim}
##  model term                   df1 df2 F.ratio p.value
##  Conclusion                     2 557 410.083  <.0001
##  Picture                        1 557   1.619  0.2038
##  Algorithm                      1 557   0.100  0.7514
##  Conclusion:Picture             2 557   0.660  0.5175
##  Conclusion:Algorithm           2 557   1.150  0.3175
##  Picture:Algorithm              1 557   0.692  0.4059
##  Conclusion:Picture:Algorithm   2 557   0.538  0.5845
\end{verbatim}

\hypertarget{binomial-responses}{%
\section{Binomial Responses}\label{binomial-responses}}

\hypertarget{do-you-think-guns-leave-unique-markings-on-discharged-bulletscasings}{%
\subsection{Do you think guns leave unique markings on discharged bullets/casings?}\label{do-you-think-guns-leave-unique-markings-on-discharged-bulletscasings}}

Responses were recorded as in a yes/no format.
16 individuals indicated that they did not think think firearms left unique markings, out of 569 total responses.

\begin{verbatim}
## Analysis of Deviance Table
## 
## Model: binomial, link: logit
## 
## Response: unique_num
## 
## Terms added sequentially (first to last)
## 
## 
##            Df Deviance Resid. Df Resid. Dev Pr(>Chi)  
## NULL                         568     145.83           
## Conclusion  2  2.12000       566     143.71   0.3465  
## Picture     1  2.88875       565     140.82   0.0892 .
## Algorithm   1  0.00591       564     140.81   0.9387  
## ---
## Signif. codes:  0 '***' 0.001 '**' 0.01 '*' 0.05 '.' 0.1 ' ' 1
\end{verbatim}

\hypertarget{conviction}{%
\subsection{Conviction}\label{conviction}}

\begin{figure}

{\centering \includegraphics{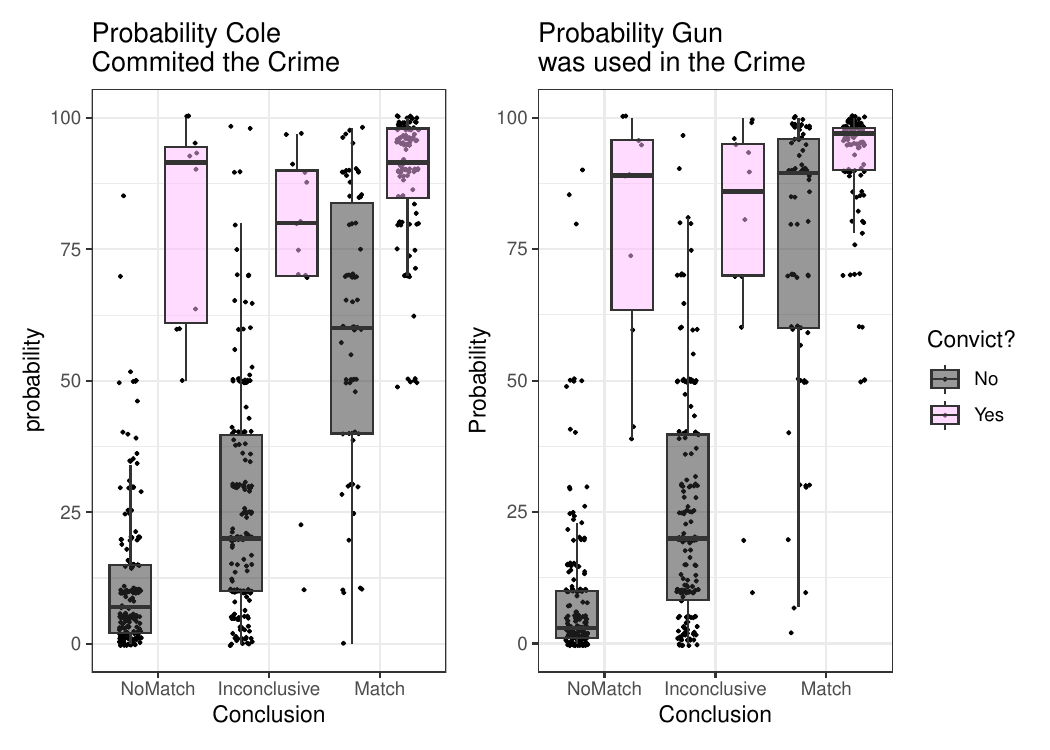} 

}

\caption{Probabilities based on whether the participants thought the defendant was guilty}\label{fig:probguilt}
\end{figure}

Individuals were given the following question:
``The State has the burden of proving beyond a reasonable doubt that the defendant is the person who committed the alleged crime.
If you are not convinced beyond a reasonable doubt that the defendant is the person who committed the alleged crime, you must find the defendant not guilty.
Would you convict this defendant, based on the evidence that you have heard?''.
Results are shown in Figure \ref{fig:probguilt}.

\begin{verbatim}
## Analysis of Deviance Table
## 
## Model: binomial, link: logit
## 
## Response: guilt_num
## 
## Terms added sequentially (first to last)
## 
## 
##                              Df Deviance Resid. Df Resid. Dev Pr(>Chi)    
## NULL                                           568     623.51             
## Conclusion                    2  207.027       566     416.48  < 2e-16 ***
## Picture                       1    3.301       565     413.18  0.06924 .  
## Algorithm                     1    3.656       564     409.53  0.05588 .  
## Conclusion:Picture            2    0.125       562     409.40  0.93927    
## Conclusion:Algorithm          2    5.088       560     404.31  0.07855 .  
## Picture:Algorithm             1    0.494       559     403.82  0.48225    
## Conclusion:Picture:Algorithm  2    4.215       557     399.60  0.12156    
## ---
## Signif. codes:  0 '***' 0.001 '**' 0.01 '*' 0.05 '.' 0.1 ' ' 1
\end{verbatim}

\bibliography{bibliography.bib}
\bibliographystyle{jds}